# Comprehensive Mapping of Tracer Diffusivities Across Composition Space in Ternary Ni–Al–Ti and Quinary Ni–Co–Fe–Al–Ti High-Entropy Alloy Using Diffusion Couple Experiments and Physics-Informed Neural Network Inversion

Ismail Kamil Worke[1], Suman Sadhu[1], Saswata Bhattacharyya[2] and Aloke Paul[1]

[1]Department of Materials Engineering, Indian Institute of Science, Bengaluru 560012, India
[2]Department of Materials Science and Metallurgical Engineering, Indian Institute of Technology, Hyderabad, Sangareddy 502284, Telangana, India
*Corresponding authors: Ismail K Worke (ismailkamil@iisc.ac.in), Saswata Bhattacharyya (saswata@msme.iith.ac.in), Aloke Paul (aloke@iisc.ac.in)

**Abstract**

A comprehensive experimental and physics-informed neural network (PINN) numerical inverse diffusion analysis is conducted in technologically important Ni-Al-Ti ternary and Ni-Co-Fe-Al-Ti quinary solid solutions for estimating and extracting composition-dependent diffusion coefficients. A systematic variation of tracer, intrinsic and interdiffusion coefficients with composition could be estimated in the ternary solid solution by intersecting a ternary (in which all the elements Ni, Ti and Al produce diffusion profiles) with three pseudo-binary diffusion profiles (in which only Ti and Al produce diffusion profiles, keeping Ni constant in different concentrations). Following, composition-dependent diffusion coefficients are extracted using the PINN method. Fick's laws, Onsager formalism, and Boltzmann scaling are embedded into the loss function in this optimization method. Self and impurity diffusion coefficients in Ni and experimentally estimated diffusion coefficients at one of the cross compositions (with PB profile) are used as equality constraints to avoid ill-posed solutions from profile-only fits. Following, the possibility of producing Al-Ti (constant Ni, Co, Fe) PB diffusion profiles in the quinary system is demonstrated. The estimation of diffusion coefficients of all the elements at the Kirkendall marker plane of a single diffusion couple profile is elaborated, which is preferably produced by coupling Ni and (Ni, Co, Fe)$_{94}$Al$_3$Ti$_3$. PINN optimisation parameters are established using self and impurity diffusion coefficients in Ni and tracer diffusion coefficients at the Kirkendall marker plane. The reliability of optimized parameters is validated by comparing with the interdiffusion coefficients estimated from binary Ni-Ti, Ni-Al and PB diffusion profiles, indicating extendibility to even lower order systems.

# 1. Introduction

Diffusion analysis, especially in multicomponent systems, is scarce because of difficulties associated with directly estimating the interdiffusion coefficients, which have been considered impossible for many decades. Only recently has this problem been solved by the innovative diffusion design couples, such as the pseudo-binary [1], pseudo-ternary [2] and body diagonal methods [3], and estimation from a single diffusion profile [4] along with the modifications of suitable equation schemes when required, correlating different types of diffusion coefficients. The concept of pseudo-binary and pseudo-ternary methods helped understand the diffusion process in solid solutions and intermetallic compounds, which is important for structural applications [5-13]. The body diagonal method is a direct extension of the conventional method for estimating interdiffusion coefficients by passing (n-1) diffusion paths (even when not intersected), preferably in a composition range in which diffusion coefficients do not vary significantly. To circumvent the problem of the requirement of ($n$-1) diffusion paths to pass reasonably closely (which is not straightforward in higher-order systems), Dash and Paul [14] proposed to estimate first the tracer diffusion coefficients from interdiffusion fluxes of only two closely passed diffusion profiles. Following, one can calculate the intrinsic and interdiffusion coefficients. This is a significant benefit since only two diffusion profiles are required, irrespective of the number of elements. Following this, Paul's group also demonstrated further ease of estimation of diffusion coefficients by intersecting or passing closely two dissimilar diffusion profiles [15-17]. The advantage of estimating data by intersecting a pseudo-binary with conventional diffusion profiles is demonstrated [18]. Very recently, the estimation of diffusion coefficients from a single diffusion profile has been demonstrated in ternary [4, 18-20] and multicomponent systems [4], which is more suitable for Ni-, Co-, Ti-, and Fe-rich alloys.

Ni-Al-Ti is one of the very important subsystems for Ni-based and high-entropy alloys. Recently, high entropy alloys in the $(NiCoFe)_x(Al, Ti)_{100-x}$ (x = atomic%) system have been proposed to have superior properties [21–32]. The diffusion process plays a crucial role in coarsening kinetics and mechanical integrity. To our knowledge, only one study is available in the Ni-Al-Ti system [33]. However, the reported interdiffusion coefficients (a kind of average of $n$ intrinsic diffusion coefficients in an $n$-component system) are not enough to highlight the diffusional interactions of a particular element, unless intrinsic diffusion coefficients are also estimated [4, 15, 17]. An element's intrinsic diffusion coefficients signify a specific element's diffusion and correctly indicate its diffusional interaction. Additionally, tracer diffusion coefficients are the fundamental diffusion parameters indicating the relative mobilities of the components, which are essential for understanding the atomic mechanism of diffusion in the absence of thermodynamic driving forces. Moreover, no diffusion studies have been reported until now in the quinary $(NiCoFe)_x(Al, Ti)_{100-x}$ high entropy alloy. An extensive diffusion study will help to quantify the diffusion-controlled processes in this crucial material system, which is already shown to possess superior mechanical properties [21, 32].

Therefore, this study aims first to estimate all types, i.e. interdiffusion, intrinsic and tracer diffusion coefficients, in the FCC phase of the ternary Ni-Al-Ti system. The method of intersecting two conventional diffusion profiles (in which all the elements produce the diffusion profiles) estimates diffusion coefficients at random compositions. Therefore, we have opted for the method of intersecting conventional diffusion profiles with pseudo-binary diffusion profiles for the estimation of data with systematic variation of composition. Since intersecting the diffusion paths in a multicomponent system is tricky, it requires very controlled experiments and multiple corrections of the end-member composition of diffusion couples. We have recently estimated the diffusion coefficients at the Kirkendall marker plane from a single diffusion profile [4], which is suitable for the composition range in the multicomponent system studied in this article. Moreover, diffusion coefficients can be determined experimentally only at a limited number of discrete compositions. Therefore, we employ a novel Physics-Informed Neural Network (PINN) numerical inverse method to extract composition-dependent diffusion coefficients as a continuous function across the entire composition range of the diffusion couple. In our approach, we use the PINN framework to embed the governing diffusion equation directly into the training process, where the network minimizes the residual of Fick's second law while matching the measured composition profiles. At the same time, equality constraints, such as tracer diffusivities at the Kirkendall marker plane and impurity diffusivities in pure Ni, are enforced to prevent ill-posed solutions and ensure that the extracted diffusion coefficients remain physically meaningful. Furthermore, the Boltzmann transformation is applied to reduce the partial differential equation of Fick's second law to an ordinary differential equation (ODE) appropriate for the annealing time of the diffusion couple. Optimizing diffusion profiles alone does not yield reliable tracer diffusivities; experimentally measured tracer diffusivities must be incorporated as equality constraints together with the diffusion profiles to obtain physically meaningful results. Accordingly, this study combines high-quality experimental data with a PINN-based numerical inverse framework to enable an in-depth diffusion analysis in ternary and multicomponent systems of major technological importance. We have demonstrated that including tracer data at intersections provides critical constraints for the numerical inverse method and ensures that the extracted diffusion coefficients correspond to the fundamental tracer diffusivities rather than being just parameters adjusted to fit the profiles. Moreover, the extendibility of optimised parameters established in a higher-order system to a lower-order system is demonstrated.

## 2. Experimental and Numerical methods for estimation and optimization

### 2.1 Experimental method

The alloy buttons were produced by melting pure elements (99.95 - 99.99 wt.%) in an arc melting unit under an argon atmosphere. Before melting, the chamber was evacuated and then filled with Ar. Each button was remelted 5-6 times for better remixing. Following, homogenization treatment of the alloy buttons was conducted at 1200$^\circ$C for 25 h in a vacuum tube furnace (~ 10$^{-4}$ Pa). After metallographic preparations, the average compositions of the alloys were measured at random spots by WDS (wavelength dispersive spectroscopy) in EPMA (electron probe micro analyzer). Pure elements were used as standards for these measurements. The Target and average compositions of the alloys prepared for different diffusion couples are listed in Table 1a. Following, thin slices of ~1.5 mm were cut using EDM (electro-discharge machine) from the homogenized buttons. After standard metallographic preparation, the diffusion couples were assembled in a fixture in the vacuum tube furnace. Annealing temperatures and times for different diffusion couples are listed in Table 1b. For detecting the Kirkendall marker plane, yttria particles with an average particle size of 1-2 μm were applied on one of the end members. After the experiment, the diffusion couples were cross-sectioned by a slow-speed diamond saw. These were then prepared metallographically for WDS line profile measurements in the EPMA. The location of the Kirkendall marker plane was identified by the presence of yttria particles along this plane. Refs. [34, 35] provide a detailed description of producing diffusion couples.

Table 1: (a) Alloy target and actual compositions used to produce the diffusion couples, (b) Diffusion couples, temperature and time of annealing.

| Alloy Designation | Target composition (at%) | Actual composition (at%) |
|---|---|---|
| A1 | $Ni_{97.5}Ti_{2.5}$ | $Ni_{97.48}Ti_{2.52}$ |
| A2 | $Ni_{95}Ti_{5}$ | $Ni_{95.05}Ti_{4.95}$ |
| A3 | $Ni_{92.5}Ti_{7.5}$ | $Ni_{92.65}Ti_{7.35}$ |
| A4 | $Ni_{97.5}Al_{2.5}$ | $Ni_{97.49}Al_{2.51}$ |
| A5 | $Ni_{95}Al_{5}$ | $Ni_{95.04}Al_{4.96}$ |
| A6 | $Ni_{92.5}Al_{7.5}$ | $Ni_{92.93}Al_{7.07}$ |
| A7 | $Ni_{90}Ti_{5}Al_{5}$ | $Ni_{89.7}Ti_{5.3}Al_{5.0}$ |
| A8 | $Ni_{31.6}Co_{31.6}Fe_{31.6}Al_{5.2}$ | $Ni_{30.9}Co_{32.1}Fe_{32.1}Al_{4.9}$ |
| A9 | $Ni_{31.6}Co_{31.6}Fe_{31.6}Ti_{5.2}$ | $Ni_{30.8}Co_{32.1}Fe_{32.2}Ti_{4.9}$ |
| A10 | $Ni_{31.3}Co_{31.3}Fe_{31.3}Ti_{3}Al_{3.1}$ | $Ni_{30.7}Co_{31.4}Fe_{32.5}Ti_{2.5}Al_{2.9}$ |

(a)

| End Members | Diffusion couples | Temperature (°C) | Time (h) |
|---|---|---|---|
| A1-A4 | DC1:PB 2.5 Ti-Al (Ni) | 1050, 1100, 1150, 1200 | 16 |
| A2-A5 | DC2: PB 5 Ti-Al (Ni) | 1050, 1100, 1150, 1200 | 16 |
| A3-A6 | DC3: PB 7.5 Ti-Al (Ni) | 1050, 1100, 1150, 1200 | 16 |
| Ni-A7 | DC4: Ternary Ni-Al-Ti | 1050, 1100, 1150, 1200 | 16 |
| A8-A9 | DC6:PB 5.5 Al-Ti (Ni-Co-Fe) | 1200 | 25 |
| Ni-A10 | DC6: Quinary Ni-Co-Fe-Al-Ti | 1200 | 25 |
| Ni-A2 | DC7: Binary Ni-Ti | 1200 | 16 |
| Ni-A5 | DC8: Binary Ni-Al | 1200 | 16 |

(b)

## 2.2 Estimation of diffusion coefficients from conventional diffusion couples

In this article, the estimations of diffusion coefficients are demonstrated using binary, pseudo-binary, ternary, and multicomponent (quinary) diffusion profiles.

In a binary system, the interdiffusion coefficient ($\widetilde{D}$) can be calculated from

$$V_m \tilde{J}_i = -\widetilde{D} \frac{\partial N_i}{\partial x} \tag{1a}$$

$$V_m \tilde{J}_i = -\frac{(N_i^+ - N_i^-)}{2t}\left[(1-Y_i^*)\int_{x^-\infty}^{x^*} Y_i dx + Y_i^* \int_{x^*}^{x^{+\infty}}(1-Y_i)dx\right] \tag{1b}$$

where $\tilde{J}_i$ is the molar interdiffusion flux (mole/m$^2$.s), $V_m$ is the molar volume (m$^3$/mol), $N_i$ is the composition in atomic/mole fraction, x is the position parameter, "*" represents the position of interest, $t$ is the diffusion annealing time of the diffusion couple, $Y_i = \frac{N_i - N_i^-}{N_i^+ - N_i^-}$ is Sauer-Freise composition a normalized variable, $N_i^-$ and $N_i^+$ are the unaffected composition of diffusion couple end members on the left- and right-hand side of the diffusion couple [34–38]. Note that we have the same interdiffusion coefficient in a binary system, irrespective of the diffusion profile of an element considered for the estimation, since we have $\tilde{J}_i + \tilde{J}_j = 0$ because of $N_i + N_j = 1$ is a binary system of elements $i$ and $j$.

In a ternary and multicomponent system, the interdiffusion coefficients are calculated from [34, 39]

$$V_m \tilde{J}_i = -\sum_{i=1}^{n-1} \tilde{D}_{ij}^n \frac{\partial N_j}{\partial x} \tag{2}$$

where $\tilde{D}_{ii}^n$ is the main interdiffusion coefficient of $i$ related to the composition gradient of the same element and $\tilde{D}_{ij}^n$ is the cross interdiffusion coefficient of the same element related to the composition gradient of another element $j$. Element $n$ is considered as the dependent variable, such that each interdiffusion flux is related to $(n-1)$ interdiffusion coefficients. Since $\sum_i^n \tilde{J}_i = 0$ because of $\sum_i^n N_i = 1$, we need to determine $(n-1)^2$ interdiffusion coefficients. Eq. 1b is applicable for calculating interdiffusion flux irrespective of the number of elements in a system.

The intrinsic fluxes are related to intrinsic diffusion coefficients in a multicomponent system by [34, 49]

$$V_m J_i = -\sum_{j=1}^{n-1} D_{ij}^n \frac{\partial N_j}{\partial x}, \tag{3a}$$

where $D_{ij}^n$ (m²/s) is the intrinsic diffusion coefficient of the element $i$ related to the composition gradient of element $j$. Therefore, the intrinsic flux of an element is related to $(n-1)$ intrinsic diffusion coefficients in which the element $n$ is considered the dependent variable. Since we have $n$ intrinsic fluxes, $n(n-1)$ intrinsic diffusion coefficients need to be estimated.

The intrinsic fluxes of elements at the Kirkendall marker plane can be calculated from the diffusion profiles developed in the interdiffusion zone from [40-42]

$$V_m J_i = -\frac{1}{2t} \left[ N_i^+ \int_{x^-}^{x_K} Y_i dx - N_i^- \int_{x_K}^{x^+} (1 - Y_i) \, dx \right], \tag{3b}$$

The intrinsic and interdiffusion coefficients are related [34,39]

$$\tilde{D}_{ij}^n = D_{ij}^n - N_i \sum_{k=1}^n D_{kj}^n \tag{4}$$

The intrinsic diffusion coefficients are related to the tracer diffusion coefficients ($D_i^*$) of elements by Manning [43], considering Onsager's cross phenomenological constants [44, 45]

$$D_{ij}^n = \frac{N_i}{N_j} D_i^* \emptyset_{ij}^n (1 + W_{ij}^n) \text{ such that } W_{ij}^n = \frac{2}{S_o \emptyset_{ij}^n} \frac{\sum_{i=1}^n N_i D_i^* \emptyset_{ij}^n}{\sum_{i=1}^n N_i D_i^*}, \tag{5}$$

where $(1 + W_{ij}^n)$ is the vacancy wind effect, a factor that originated from Onsager's cross terms, $S_o$ is a structure factor equal to 7.15 for FCC alloys, as considered in this study [43,46] and $\emptyset_{ij}^n$ is the thermodynamic factor related to the activity gradient by $\emptyset_{ij}^n = \frac{\partial ln a_i}{\partial ln N_1} - \frac{N_1}{N_n} \frac{\partial ln a_i}{\partial ln N_n}$.

## 2.3 Estimation of diffusion coefficients from binary and constrained pseudo-binary (PB) diffusion couples

In a *i-j* binary system, the composition-dependent interdiffusion coefficients can be estimated from [34]

$$V_m \tilde{J}_{i(or\,j)} = -\tilde{D} \frac{\partial N_{i(or\,j)}}{\partial x}, \tag{6a}$$

The interdiffusion flux can be calculated from Eq. 1b. The interdiffusion and intrinsic diffusion coefficients are related by [34, 47]

$$\tilde{D} = N_j D_i + N_i D_j \tag{6b}$$

The intrinsic and tracer diffusion coefficients are related by [46]

$$D_i = D_i^* \left[ 1 + \frac{2N_i(D_i^* - D_j^*)}{S_o(N_i D_i^* + N_j D_j^*)} \right] = D_i^*(1 + W_i), \tag{6c}$$

$$D_j = D_j^* \left[ 1 - \frac{2N_j(D_i^* - D_j^*)}{S_o(N_i D_i^* + N_j D_j^*)} \right] = D_j^*(1 - W_j), \tag{6d}$$

where $(1 + W_i)$ and $(1 - W_j)$ are the vacancy wind effects of element *i* and *j*.

There are similarities in the equation scheme in the pseudo-binary diffusion couple with the binary diffusion couple in which only elements, let's say *i* and *j* develop the diffusion profiles, keeping the other elements constant. The reasoning behind the equation scheme established for the PB diffusion couple can be found elsewhere [48]. The issues with PB non-ideal diffusion profiles are also discussed extensively when certain elements are attempted to keep constant but produce the diffusion profiles [48]. Such issues are not faced with the PB diffusion couples produced in the ternary and multicomponent systems produced here. Therefore, the equations scheme related to ideal/near-ideal diffusion profiles is applicable. In this, the interdiffusion flux and the PB interdiffusion coefficient are related by [48].

$$V_m \tilde{J}_i = -\widetilde{D}_{PB} \frac{\partial N_i}{\partial x} \tag{7a}$$

One can also calculate this directly by dividing the composition of two diffusing elements $i$ and $j$ i.e. $(N_i + N_j)$ on both sides of Eq. 1b and 6a to have

$$\frac{V_m \tilde{J}_i}{(N_i + N_j)} = -\frac{(M_i^+ - M_i^-)}{2t}\left[(1 - Y_{i,M}^*)\int_{x-\infty}^{x^*} Y_{i,M} dx + Y_{i,M}^* \int_{x^*}^{x+\infty}(1 - Y_{i,M})dx\right], \tag{7b}$$

$$\frac{V_m \tilde{J}_i}{(N_i + N_j)} = -\widetilde{D}_{PB}\frac{\partial M_i}{\partial x}, \tag{7c}$$

where $M_i = \frac{N_i}{N_i + N_j}$ and $Y_{i,M} = \frac{M_i - M_i^-}{M_i^+ - M_i^-} = \frac{N_i - N_i^-}{N_i^+ - N_i^-} = Y_i$.

Correlating these two equations, we can directly estimate the PB interdiffusion coefficient from the diffusion profile by

$$\widetilde{D}_{PB}\frac{\partial M_i}{\partial x} = \frac{(M_i^+ - M_i^-)}{2t}\left[(1 - Y_{i,M}^*)\int_{x-\infty}^{x^*} Y_{i,M} dx + Y_{i,M}^* \int_{x^*}^{x+\infty}(1 - Y_{i,M})dx\right], \tag{7d}$$

The PB interdiffusion coefficient is related to the intrinsic diffusion coefficients by [48]

$$\widetilde{D}_{PB} = M_j D_{ii}^j + M_i D_{jj}^i \tag{8a}$$

The intrinsic diffusion coefficients are related to the tracer diffusion coefficients by [48]

$$D_{ii}^j = D_i^* \phi_{ii}^j \left[1 + \frac{2N_i\left(D_i^* - D_j^*\frac{\phi_{jj}^i}{\phi_{ii}^j}\right)}{S_o(N_i D_i^* + N_j D_j^*)}\right] = D_i^* \phi_{ii}^j (1 + W_{ii}^{j,PB}), \tag{8b}$$

$$D_{jj}^i = D_j^* \phi_{jj}^i \left[1 - \frac{2N_j\left(D_i^*\frac{\phi_{ii}^j}{\phi_{jj}^i} - D_j^*\right)}{S_o(N_i D_i^* + N_j D_j^*)}\right] = D_j^* \phi_{jj}^i (1 - W_{jj}^{i,PB}), \tag{8c}$$

where $(1 + W_{ii}^{j,PB})$ and $(1 - W_{jj}^{i,PB})$ are the vacancy wind effects of the elements in a PB diffusion couple.

## 2.5 PINN-based numerical method for optimization

A physics-informed neural network (PINN) inverse method, implemented with the open-source DeepXDE library [49], is used to infer composition-dependent diffusivities from measured diffusion profiles and independently estimated tracer/self/impurity diffusivities at selected compositions. Using profile data and pointwise constraints in a single objective improves identifiability and suppresses non-physical solutions that might otherwise fit the profiles alone.

To maintain a numerically well-scaled problem, tracer diffusivities are nondimensionalized and optimised in log space. We define [50]

$$\bar{D}_i^* = \frac{D_i^*}{D_0}, \tag{9a}$$

so that

$$D_i^* = D_0 \bar{D}_i^*. \tag{9b}$$

We train the network on $\log \bar{D}_i^*$ rather than on $D_i^*$ directly. The reference $D_0$ is any fixed positive constant; it sets only the numerical scale and does not change the recovered physics, because the learned $\bar{D}_i^*$ rescales so that $D_i^* = D_0 \bar{D}_i^*$ matches the data. Consistency is maintained by applying the same scaling to the equality-constraint values, $\bar{D}_i^{\text{eq}} = D_i^{\text{eq}}/D_0$. In the Manning vacancy–wind factor, this scaling cancels:

$$\xi = \frac{2}{S_0 \sum_m N_m D_m^*} = \frac{1}{D_0} \frac{2}{S_0 \sum_m N_m \bar{D}_m^*},$$

so $D_0$ does not alter the vacancy–wind contribution. Keeping $D_0$ fixed within each run stabilizes the loss landscape and improves comparability across experiments.

The composition dependence of the (non-dimensional) tracer diffusivity is expressed in log-space (with the $n$-th component eliminated by $\sum_{m=1}^{n} N_m = 1$) as

$$\log(\bar{D}_i^*) = \theta_0^{(i)} + \sum_{k=1}^{n-1} \theta_k^{(1,i)} N_k + \sum_{j=1}^{n-2} \sum_{k=j+1}^{n-1} \theta_{jk}^{(2,i)} N_j N_k. \tag{10}$$

$\Theta_i = \{\theta_0^{(i)}, \theta_k^{(1,i)}, \theta_{jk}^{(2,i)}\}$ are trainable, dimensionless parameters; the inputs $N_k$ are mole fractions (dimensionless), and $\log \bar{D}_i^*$ is the log of a dimensionless ratio. Working in log-space guarantees positivity of $D_i^*$ and converts multiplicative variability into an additive structure that is better conditioned for gradient-based optimization.

To connect tracer diffusivities to mass transport, we write the intrinsic fluxes in the Onsager form~[30,34]:

$$J_i = -\sum_{k=1}^{n} L_{ik} \frac{\partial \mu_k}{\partial x}, \quad (11)$$

with phenomenological coefficients $L_{ik}$. Following Manning's vacancy–wind correlation in the tracer-diffusivity convention~[44],

$$L_{ii} = \frac{C_i D_i^*}{RT}(1 + \xi\, N_i D_i^*),\; L_{is} = \frac{C_s D_s^*}{RT}(\xi\, N_i D_i^*)\; (i \neq s), \quad (12)$$

where $S_0 = 7.15$ for FCC and $\xi = 2/(S_0 \sum_m N_m D_m^*)$. Substituting Eq. 12 into Eq. 11 and multiplying by the molar volume $V_m$ gives

$$V_m J_i = -\frac{N_i D_i^*}{RT} \frac{\partial \mu_i}{\partial x} - \frac{\xi N_i D_i^*}{RT} \sum_{k=1}^{n} N_k D_k^* \frac{\partial \mu_k}{\partial x}. \quad (13)$$

The interdiffusion flux is then

$$\tilde{J}_i = J_i - N_i \sum_{k=1}^{n} J_k, \quad (14)$$

and, assuming constant molar volume, Fick's second law [51, 52] relates $\tilde{J}_i$ to the composition field [34, 39]:

$$\frac{\partial N_i}{\partial t} = -\frac{\partial}{\partial x}(V_m \tilde{J}_i). \quad (15)$$

Because experiments typically provide a single profile at a fixed anneal time, it is convenient to reduce the PDE to an ODE using the classical Boltzmann variable $\lambda = x/\sqrt{t}$

$$-\frac{\lambda}{2} \frac{dN_i}{d\lambda} + \frac{d}{d\lambda}[V_m \tilde{J}_i(\lambda)] = 0. \quad (16)$$

For numerical stability, we normalize $\lambda$ to $\bar{\lambda} \in [0,1]$ via $\bar{\lambda} = (\lambda - \lambda^-)/(\lambda^+ - \lambda^-)$ and define $\eta = t/(x^+)^2$, which yields

$$-\frac{\bar{\lambda}}{2}\frac{dN_i}{d\bar{\lambda}} + \eta \frac{d}{d\bar{\lambda}}[V_m \tilde{J}_i(\bar{\lambda})] = 0, \lambda^-, \lambda^+: \text{unaffected end-member regions}. \tag{17}$$

Eq. [17] is the residual enforced by the network at collocation points.

The inverse problem is posed by seeking a profile $N_i(\bar{\lambda})$ and parameters $\Theta$ that satisfy the physics while matching both the measured profile and diffusivity constraints. In compact form, the conditions read

$$\begin{aligned}
&\text{PDE residual:} & -\frac{\bar{\lambda}}{2}\frac{dN_i}{d\bar{\lambda}} + \eta \frac{d}{d\bar{\lambda}}[V_m \tilde{J}_i(\bar{\lambda})] &= 0, \\
&\text{Boundary values (endpoints in Boltzmann space):} & N_i(0) = N_i^L, N_i(1) &= N_i^R, \\
&\text{Equality constraints at selected compositions } N_u: & D_i^*(N_u) &= D_i^{eq}(N_u).
\end{aligned} \tag{18}$$

All equality constraints are applied in the same scaled form, replacing $D_i^*$ by $D_0 \bar{D}_i^*$ and $D_i^{eq}$ by $\bar{D}_i^{eq} = D_i^{eq}/D_0$; as noted above, this re-scaling leaves the physics unchanged but balances term magnitudes in the loss.

The objective minimized by the PINN has four components,

$$L_{\text{total}} = \omega_r L_r + \omega_b L_b + \omega_d L_d + \omega_c L_c, \tag{19}$$

with

$$L_r = \frac{1}{P_r}\sum_{p=1}^{P_r}\left(-\frac{\bar{\lambda}_p}{2}\frac{dN_i}{d\bar{\lambda}} + \eta \frac{d}{d\bar{\lambda}}(V_m \tilde{J}_i)\right)^2_{\bar{\lambda}=\bar{\lambda}_p}, \tag{20a}$$

$$L_b = \frac{1}{P_b}\sum_{q=1}^{P_b}[(N_i(0) - N_i^L)^2 + (N_i(1) - N_i^R)^2], \tag{20b}$$

$$L_d = \frac{1}{P_d}\sum_{p=1}^{P_d}\|N_i(\bar{\lambda}_p) - N_i^{\exp}(\bar{\lambda}_p)\|^2, \tag{20c}$$

$$L_c = \frac{1}{P_c}\sum_{u=1}^{P_c}\|D_i^{\text{model}}(N_u) - D_i^{eq}(N_u)\|^2. \tag{20d}$$

Here $P_r, P_b, P_d$, and $P_c$ are the numbers of collocation points for the residual, boundary, data, and equality-constraint terms, and $(\omega_r, \omega_b, \omega_d, \omega_c)$ weight their relative influence. Gradients are obtained by automatic differentiation (AD), which handles the implicit dependence of $N_i$ on Θ through Eq. 17. Writing the ODE constraint abstractly as $j(\Theta, N_i) = 0$, the total derivative of the objective with respect to $\theta_j \in \Theta$ is

$$\frac{d\hat{L}(\Theta)}{d\theta_j} = \nabla_{\theta_j} L(\Theta, N_i(\Theta)) + \nabla_{N_i} L(\Theta, N_i(\Theta)) \frac{dN_i(\Theta)}{d\theta_j}, j = 1, 2, \ldots \quad (21a)$$

while the constraint sensitivity satisfies

$$\nabla_{\theta_j} j + \nabla_{N_i} j \frac{dN_i(\Theta)}{d\theta_j} = 0, \quad (21b)$$

which determines the implicit term $dN_i/d\theta_j$ used by AD. In practice, we use a first-order optimizer (with learning-rate warm-up and decay) followed by L–BFGS, and the same fixed $D_0$ is used throughout each run to maintain a stable and well-scaled optimization landscape.

## 3. Results and discussion

### 3.1 Experimentally estimated diffusion coefficients

The pseudo-binary diffusion couples are first produced in the ternary Ni-Ti-Al system by coupling Ni2.5Al-Ni2.5Ti, Ni5Al-Ni5Ti and Ni7.5Al-Ni7.5Ti alloys at 1200°C by annealing for 16 h. The diffusion paths on the Gibbs triangle are shown in Fig. 1a and the diffusion profiles produced are shown in Fig. 1b-d. Ni remained constant in all the PB diffusion profiles without showing any non-ideality beyond the range of scattered data. Therefore, these profiles can be considered as ideal/near-ideal diffusion profiles. A detailed explanation of ideal/near-ideal and non-ideal PB diffusion profiles can be found in Ref. [48]. Previously, the Ni-constant PB diffusion profile was reported in the γ'-Ni$_3$(Al, Ti) phase [33], although the correlations between the PB interdiffusion coefficients with intrinsic and tracer diffusion coefficients were not known at that time. In this study, we have shown the possibility of producing Ni-constant PB diffusion profiles in the γ-Ni(Al, Ti) solid solution phase. These experiments were conducted at four different temperatures, 1050, 1100, 1150 and 1200ºC. One can estimate the PB interdiffusion coefficients from these profiles at different temperatures. Such a calculation example is shown in the supplementary file for the 5Al-5Ti (95 at.%Ni constant). To calculate the tracer diffusion coefficients by intersecting diffusion profiles, a ternary diffusion profile was produced, as shown in Fig. 2a. This intersects all the PB profiles, as shown on the Gibbs triangle in Fig. 2b, highlighting the intersecting compositions. The intersecting compositions at different temperatures are listed in Table 2. It can be seen that these compositions are similar, facilitating the estimation of data at different temperatures for the calculation of the activation energy.

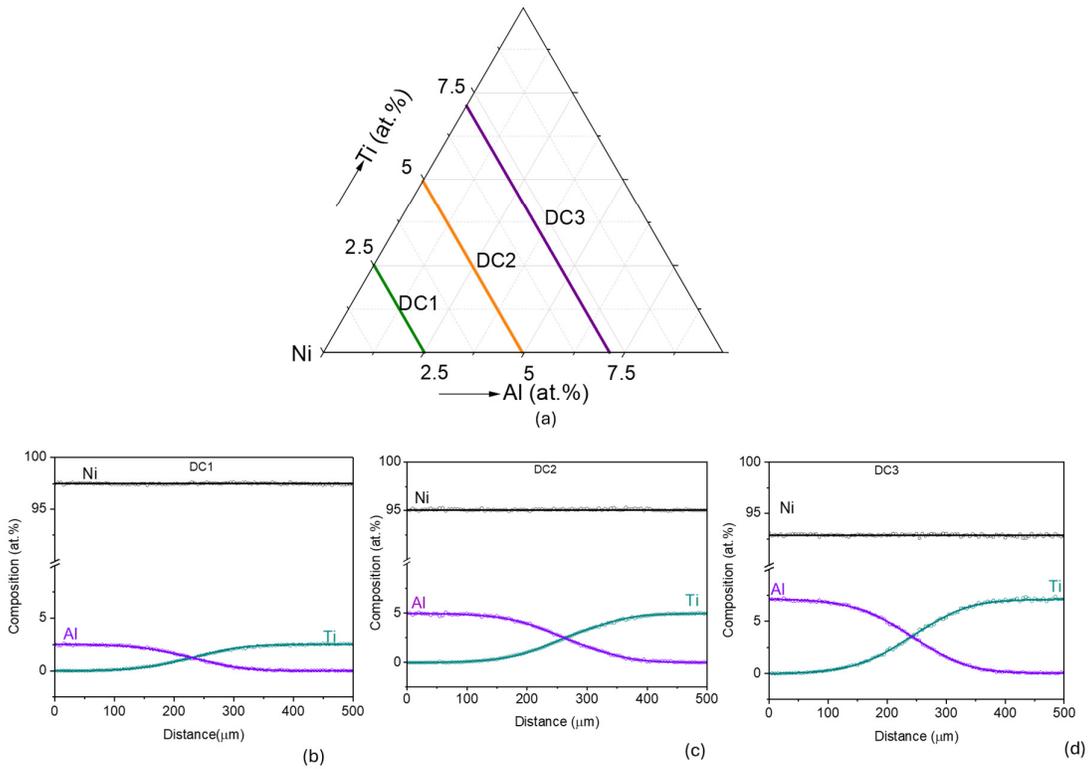

Fig.1 (a) The PB diffusion profiles produced at 1200°C after annealing for 16 h , (b) 2.5Ti-2.5Al (97.5 at.% Ni constant), (c) 5Ti-5Al (95 at.% Ni constant), (d) 7.5Ti-7.5Al (92.5 at.% Ni constant) diffusion profiles.

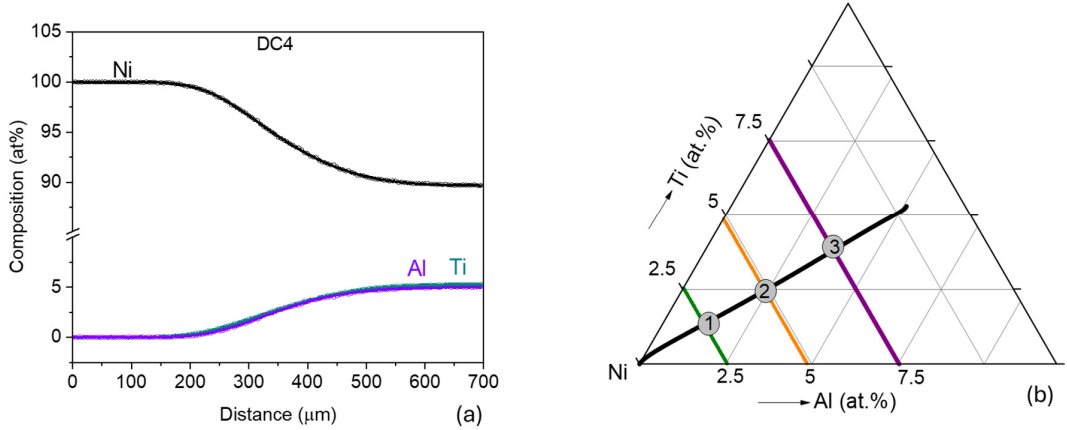

Fig. 2 (a) A ternary diffusion profiles produced at 1200°C after annealing for 16 are shown on Gibbs triangle and (b) intersecting composition with PB profiles on Gibbs triangle.

Table 2 Intersecting compositions between the ternary diffusion profile with 97.5 at.% constant PB (cross-1), 95.0 at.% constant PB (cross-2) and 92.5 at.% constant PB (cross-3) at different temperatures (1050, 1100, 1150 and 1200°C).

| Cross-1 | 1050°C (at.%) | 1100°C (at.%) | 1150°C (at.%) | 1200°C (at.%) |
|---|---|---|---|---|
| $N_{Ni}$ | 97.4 | 97.5 | 97.5 | 97.5 |
| $N_{Ti}$ | 1.4 | 1.4 | 1.4 | 1.4 |
| $N_{Al}$ | 1.2 | 1.1 | 1.1 | 1.1 |
| Cross-2 | | | | |
| $N_{Ni}$ | 95.1 | 95.0 | 95.1 | 95.1 |
| $N_{Ti}$ | 2.5 | 2.7 | 2.6 | 2.5 |
| $N_{Al}$ | 2.4 | 2.3 | 2.3 | 2.4 |
| Cross-3 | | | | |
| $N_{Ni}$ | 92.5 | 92.5 | 92.5 | 92.6 |
| $N_{Ti}$ | 3.8 | 3.9 | 3.8 | 3.7 |
| $N_{Al}$ | 3.7 | 3.6 | 3.7 | 3.7 |

At these intersecting compositions, we can express two interdiffusion fluxes from the ternary diffusion profile (DC4) with tracer diffusion coefficients of Ni, Al and Ti (which produce the diffusion profiles) by substituting Eq. 4 and 5 in Eq. 2. Additionally, at the same composition, one equation can be written from the PB diffusion couple (DC1, DC2 or DC3) expressing the interdiffusion flux with the tracer diffusion coefficients of Al and Ti (which produce the diffusion profiles) by replacing Eq. 7 and 8 in Eq. 7. Therefore, we have three equations relating three tracer diffusion coefficients, which can be calculated from the interdiffusion fluxes and composition gradients at the composition of the intersection [18]. The diffusion of Al and Ti controls the interdiffusion process in this Ni-rich alloy, and we need to estimate the tracer diffusion coefficients of these two elements. Therefore, estimation of Ni tracer diffusion coefficients can induce very high error, which is explained in detail in Ref. [48]. Thus, the tracer diffusion coefficients of Ni in pure Ni available in the literature [53] at different temperatures as $1.1 \times 10^{-15}$, $2.9 \times 10^{-15}$, $6.8 \times 10^{-15}$, $15.3 \times 10^{-15}$ at 1050, 1100, 1150 and 1200ºC, respectively, are considered for these calculations. The thermodynamic factors are calculated from the activity data available in ThermoCalc TCNI9 [54]. The estimated tracer diffusion coefficients of Ti and Al at different cross compositions and temperatures are listed in Table 3. Following, we can calculate the intrinsic diffusion coefficients from Eq. 5 and interdiffusion coefficients from Eq. 4. For example, the computed values at 1200ºC at three intersecting compositions are listed in Table 4. These data at other temperatures are given in the supplementary file. As already mentioned, Karunaratne and Reed [33] estimated interdiffusion coefficients directly by intersecting ternary diffusion profiles. The values and signs of the interdiffusion coefficients are similar between these two studies. However, systematic composition-dependent diffusivities could be estimated in this study by intersecting ternary profiles with PB profiles instead of only ternary diffusion profiles by Karunaratne and Reed. Moreover, the intrinsic and tracer diffusion coefficients are also calculated in this study.

Table 3: Tracer diffusion coefficients of Al and Ti estimated at different cross compositions at four different temperatures. The tracer diffusion coefficient of Ni in pure Ni extracted from Ref. [53] as $1.1 \times 10^{-15}$, $2.9 \times 10^{-15}$, $6.8 \times 10^{-15}$, $15.3 \times 10^{-15}$ $m^2/s$ at 1050, 1100, 1150 and 1200ºC, respectively, considered for these calculations.

| Cross-1 | 1050ºC (x10⁻¹⁵ m²/s) | 1100ºC (x10⁻¹⁵ m²/s) | 1150ºC (x10⁻¹⁵ m²/s) | 1200ºC (x10⁻¹⁵ m²/s) |
|---|---|---|---|---|
| $D^*_{Ti}$ | 5.2±0.8 | 10.1±1.5 | 22.1±3.3 | 40.1±6 |
| $D^*_{Al}$ | 5.3±0.8 | 10.5±1.6 | 24.3±3.6 | 42.7±6.5 |
| Cross-2 | | | | |
| $D^*_{Ti}$ | 4.2±0.7 | 9.0±1.4 | 21.4±3.15 | 43.4±6.5 |
| $D^*_{Al}$ | 5.6±0.75 | 11.3±1.6 | 27.2±4.1 | 57.2±8.6 |
| Cross-3 | | | | |
| $D^*_{Ti}$ | 4.0±0.7 | 9.0±1.4 | 19.1±2.9 | 41.0±6.1 |
| $D^*_{Al}$ | 5.7±0.75 | 12.1±1.6 | 27.4±4.0 | 58.9±8.9 |

It can be noted from Table 3 that the tracer diffusion coefficients of Al are a bit higher than the tracer diffusion coefficients of Ti (although within the range of uncertainty). However, because of the role of thermodynamic driving forces accounted for by the thermodynamic factors (as listed in Table 4), the main intrinsic and interdiffusion coefficients of Ti are a bit higher than the similar diffusion parameters of Al. These are listed for all three cross compositions at 1200 °C. Data at other temperatures can be found in the supplementary file. Since the diffusion coefficients are estimated in alloys with relatively small compositions of alloying Ti and Al, the intrinsic and interdiffusion coefficients of these elements are found to be similar, which can be understood from the relation Eq. 4 for a relatively small value of $N_i$ ($i$ = Al and Ti). These could differ in an alloy with a reasonably high concentration of alloying elements or high entropy alloys [15, 17, 50]. Moreover, because of the same reason, the role of the vacancy wind effects on the intrinsic diffusion coefficients of these elements is not significant. On the other hand, the cross-diffusion coefficients of Ni with these elements ($D_{NiTi}^{Ni}$ and $D_{NiAl}^{Ni}$) are influenced by the vacancy wind effect (*i.e.* the role of Onsager's cross phenomenological constants) because of the relatively higher composition of Ni, as indicated by the deviation of $(1 + W_i)$ from one. The positive values of the cross-intrinsic diffusion coefficients of Ti and Al ($D_{TiAl}^{Ni}$ and $D_{AlTi}^{Ni}$) indicate positive diffusional interactions, i.e., an increase in flux when these are diffused in the same direction with similar signs of composition gradient (similar to the diffusion direction of elements in the ternary diffusion profile produced in this study, as shown in Fig. 2a). The variation of intrinsic diffusion coefficients at three cross compositions at different temperatures are shown in Fig. 3 shown on the Gibbs triangle at different temperatures, indicating a systematic variation with composition and temperature. It is clear that all the intrinsic diffusion coefficients (main and cross) increase with the simultaneous increase in Ti and Al content (i.e. decrease in Ni-content). The calculation of the pseudo-binary interdiffusion coefficients of Ti and Al, estimated (for example, from DC2, as shown in the supplementary file), does not show much variation along Al or Ti composition variation, indicating the change in diffusion coefficients mainly with the change in Ni-content. The Arrhenius plots of tracer, main intrinsic and main interdiffusion coefficients at cross 2 composition (refer to Table 2) are shown in Fig. 4. Although there is a difference in main intrinsic (or interdiffusion coefficients) with tracer diffusion coefficients, the activation energy for all types of diffusion coefficients is found to be very similar with the value of 255±5 KJ/mole. Similar values are also estimated at other cross compositions as well. As already mentioned, the interdiffusion and intrinsic diffusion coefficients are expected to be similar since these are calculated at a relatively low composition of Ti and Al.

Table 4: Calculation of intrinsic, interdiffusion coefficients and the vacancy wind effects at 1200ºC at (a) cross composition 1, (b) cross composition 2 and (c) cross composition 3 (refer Table 2) from the estimated tracer diffusion coefficients (refer Table 3). The tracer diffusion coefficient of Ni in pure Ni extracted from Ref. [53] as $15.3 \times 10^{-15}\ m^2/s$ at 1200ºC is considered for these calculations. Thermodynamic factors are calculated from the thermodynamic data available in ThermoCalc database TCNI9 [54].

| $\Phi_{ij}^n$ | | $D_{ij}^n$ | Without VWE (x10⁻¹⁵m²/s) | With VWE (x10⁻¹⁵m²/s) | 1+$W_i$ | $\widetilde{D}_{ij}^n$ | Without VWE (x10⁻¹⁵m²/s) | With VWE (x10⁻¹⁵m²/s) | 1+$W_i$ |
|---|---|---|---|---|---|---|---|---|---|
| $\Phi_{AlAl}^{Ni}$ | 1.02 | $D_{AlAl}^{Ni}$ | 43.5 | 43.6 | 1.00 | $\widetilde{D}_{AlAl}^{Ni}$ | 43.2 | 43.2 | 1.00 |
| $\Phi_{AlTi}^{Ni}$ | 0.18 | $D_{AlTi}^{Ni}$ | 6.04 | 6.1 | 1.01 | $\widetilde{D}_{AlTi}^{Ni}$ | 5.73 | 5.77 | 1.01 |
| $\Phi_{TiAl}^{Ni}$ | 0.21 | $D_{TiAl}^{Ni}$ | 10.7 | 10.8 | 1.01 | $\widetilde{D}_{TiAl}^{Ni}$ | 9.9 | 10.0 | 1.01 |
| $\Phi_{TiTi}^{Ni}$ | 1.52 | $D_{TiTi}^{Ni}$ | 61.0 | 61.0 | 1.00 | $\widetilde{D}_{TiTi}^{Ni}$ | 60.3 | 60.4 | 1.00 |
| $\Phi_{NiAl}^{Ni}$ | -0.014 | $D_{NiAl}^{Ni}$ | -19.0 | -17.5 | 0.92 | | | | |
| $\Phi_{NiTi}^{Ni}$ | -0.02 | $D_{NiTi}^{Ni}$ | -21.3 | -19.4 | 0.91 | | | | |

(a)

| $\Phi_{ij}^n$ | | $D_{ij}^n$ | Without VWE (x10⁻¹⁵m²/s) | With VWE (x10⁻¹⁵m²/s) | 1+$W_i$ | $\widetilde{D}_{ij}^n$ | Without VWE (x10⁻¹⁵m²/s) | With VWE (x10⁻¹⁵m²/s) | 1+$W_i$ |
|---|---|---|---|---|---|---|---|---|---|
| $\Phi_{AlAl}^{Ni}$ | 1.08 | $D_{AlAl}^{Ni}$ | 61.8 | 61.9 | 1.00 | $\widetilde{D}_{AlAl}^{Ni}$ | 60.4 | 60.5 | 1.00 |
| $\Phi_{AlTi}^{Ni}$ | 0.38 | $D_{AlTi}^{Ni}$ | 20.9 | 21.1 | 1.01 | $\widetilde{D}_{AlTi}^{Ni}$ | 19.7 | 19.9 | 1.01 |
| $\Phi_{TiAl}^{Ni}$ | 0.39 | $D_{TiAl}^{Ni}$ | 17.6 | 17.8 | 1.01 | $\widetilde{D}_{TiAl}^{Ni}$ | 15.6 | 15.7 | 1.01 |
| $\Phi_{TiTi}^{Ni}$ | 1.91 | $D_{TiTi}^{Ni}$ | 82.9 | 83.1 | 1.00 | $\widetilde{D}_{TiTi}^{Ni}$ | 81.2 | 81.3 | 1.00 |
| $\Phi_{NiAl}^{Ni}$ | -0.034 | $D_{NiAl}^{Ni}$ | -20.6 | -18.5 | 0.89 | | | | |
| $\Phi_{NiTi}^{Ni}$ | -0.061 | $D_{NiTi}^{Ni}$ | -35.5 | -33.1 | 0.93 | | | | |

(b)

| $\Phi_{ij}^n$ | | $D_{ij}^n$ | Without VWE (x10⁻¹⁵m²/s) | With VWE (x10⁻¹⁵m²/s) | 1+$W_i$ | $\widetilde{D}_{ij}^n$ | Without VWE (x10⁻¹⁵m²/s) | With VWE (x10⁻¹⁵m²/s) | 1+$W_i$ |
|---|---|---|---|---|---|---|---|---|---|
| $\Phi_{AlAl}^{Ni}$ | 1.2 | $D_{AlAl}^{Ni}$ | 70.7 | 71.0 | 1.004 | $\widetilde{D}_{AlAl}^{Ni}$ | 68.2 | 68.5 | 1.00 |
| $\Phi_{AlTi}^{Ni}$ | 0.59 | $D_{AlTi}^{Ni}$ | 35.3 | 35.8 | 1.01 | $\widetilde{D}_{AlTi}^{Ni}$ | 33.0 | 33.3 | 1.01 |
| $\Phi_{TiAl}^{Ni}$ | 0.56 | $D_{TiAl}^{Ni}$ | 22.8 | 23.1 | 1.01 | $\widetilde{D}_{TiAl}^{Ni}$ | 19.5 | 19.7 | 1.01 |
| $\Phi_{TiTi}^{Ni}$ | 2.29 | $D_{TiTi}^{Ni}$ | 93.4 | 93.7 | 1.00 | $\widetilde{D}_{TiTi}^{Ni}$ | 90.2 | 90.4 | 1.00 |
| $\Phi_{NiAl}^{Ni}$ | -0.07 | $D_{NiAl}^{Ni}$ | -26.8 | -24.5 | 0.91 | | | | |
| $\Phi_{NiTi}^{Ni}$ | -0.11 | $D_{NiTi}^{Ni}$ | -42.1 | -39.1 | 0.93 | | | | |

(c)

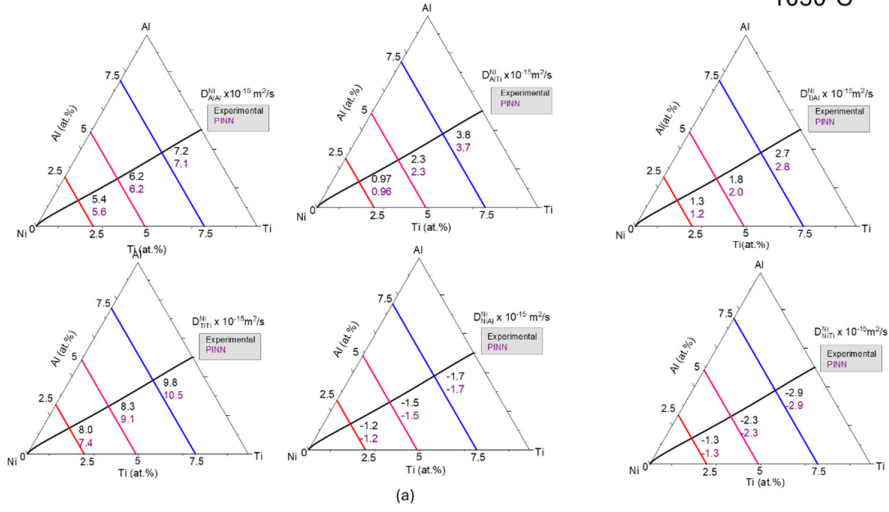

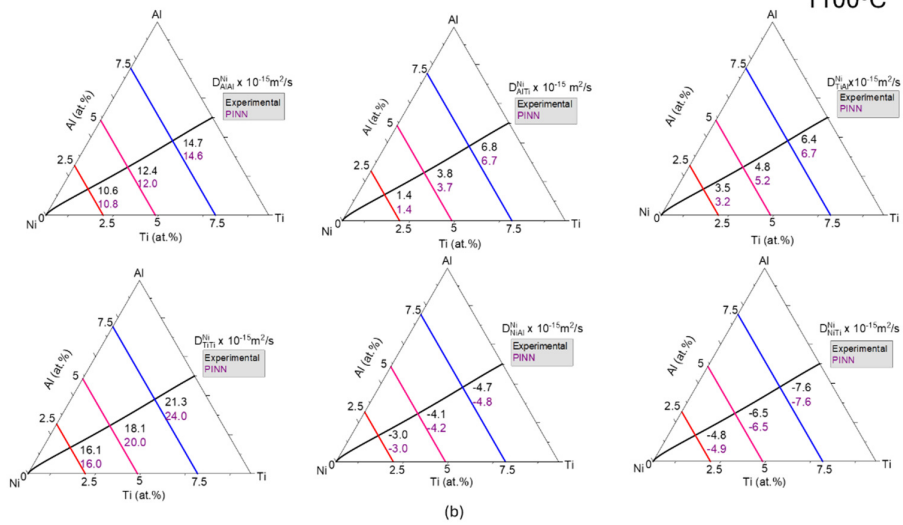

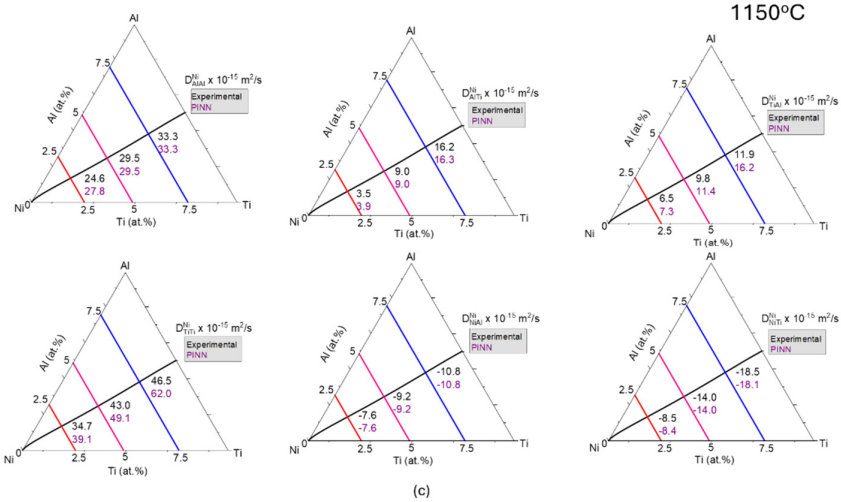

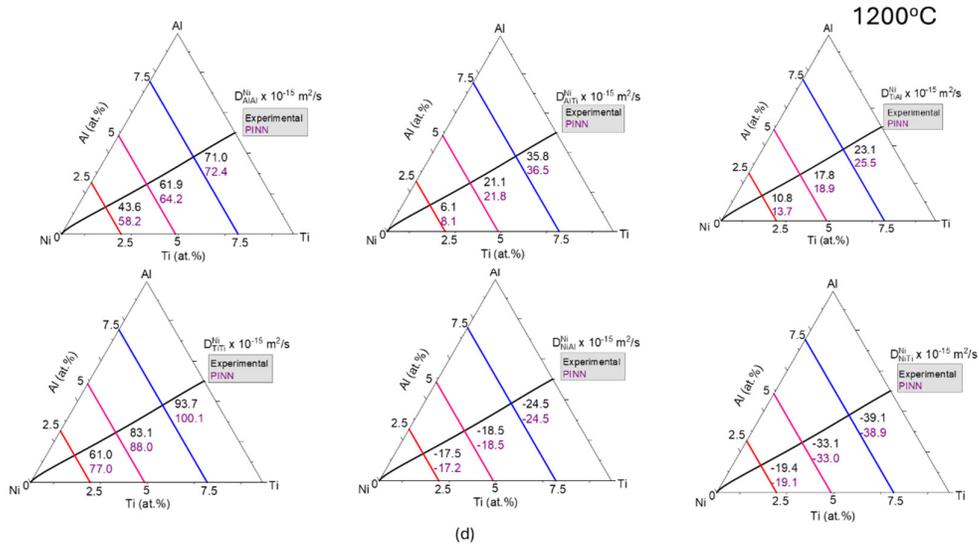

Fig. 3 Intrinsic diffusion coefficients calculated from estimated tracer diffusion coefficients at the cross compositions (refer to Table 2) in comparison to calculated data after PINN optimization at (a) 1050°C, (b) 1100°C, (c) 1150°C and (d) 1200°C.

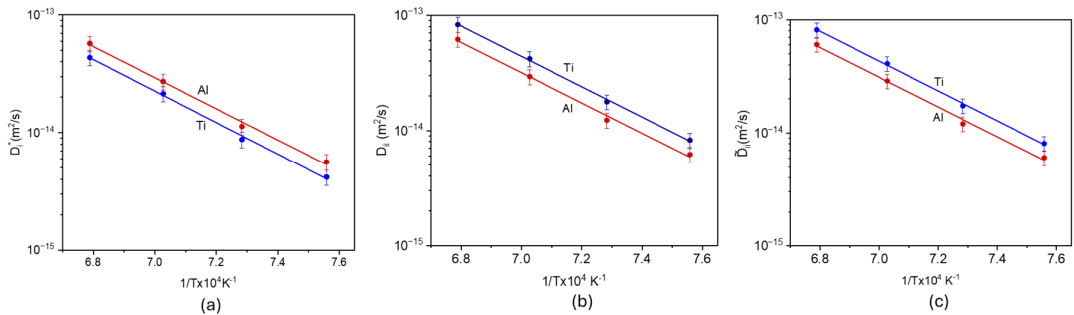

Fig. 4 Arrhenius plot of (a) Tracer, (b) Intrinsic, and (c) Interdiffusion coefficients at composition corresponding to cross 2 (see Table 2).

Following this, we aimed to estimate the diffusion coefficients in the Ni-Co-Fe-Cr-Al system. First, to check the possibilities of producing the PB diffusion profiles in this five-component system, one PB diffusion couples were produced, keeping Ni constant (95 at.%), as shown in Fig. 5a. This also produces an ideal/near-ideal PB diffusion profile without any non-ideality beyond the scatter in measured composition, similar to the ternary system. The calculated Al-Ti PB interdiffusion coefficient in this quinary system is a little higher than the interdiffusion coefficient in the ternary system, as shown in Fig. 5b. There are two possibilities to estimate the diffusion coefficients of all the elements controlling the interdiffusion process. A five-component conventional diffusion profile can be intersected exactly by a pseudo-binary diffusion profile to calculate the elements' tracer diffusion coefficients, as demonstrated in the Ni-Co-Fe-Cr system [15]. During the experimental analysis of this system, another method of estimation at the Kirkendall marker plane from a single profile was demonstrated successfully in ternary [4, 18, 19, 20], quaternary and quinary systems [4]. This method generates a mobility database over almost the whole solid solution phase of NiCoFeCrMn [50]. This is relatively easier than the previous method. Therefore, it gives an opportunity to practice this method in this system of practical application for generating a mobility database following PINN optimization. The PB diffusion couple produced can be utilized for validating optimization parameters.

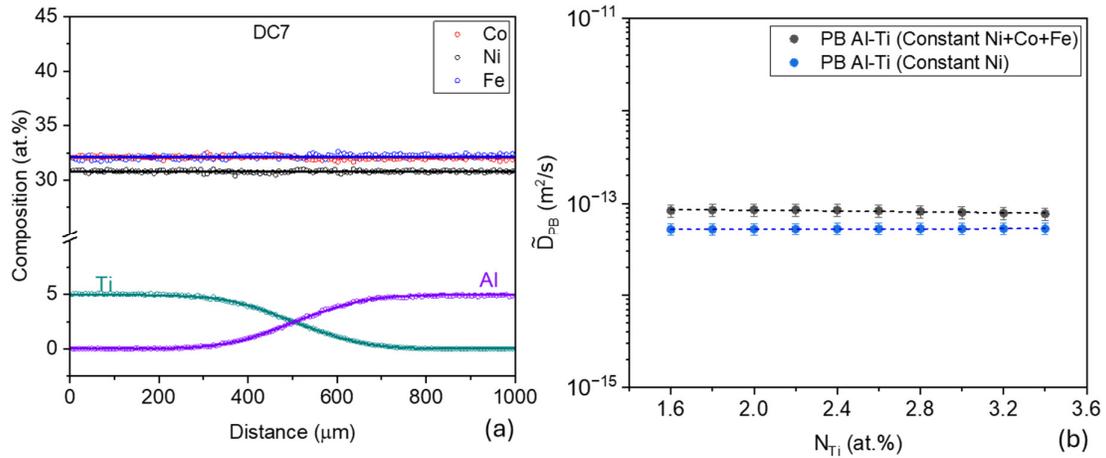

Fig. 5 (a) The PB Al-Ti (constant Ni, Co, Fe) diffusion profiles produced at 1200 ºC after annealing for 25 h and (b) PB Al-Ti interdiffusion coefficients compared between ternary and quaternary systems

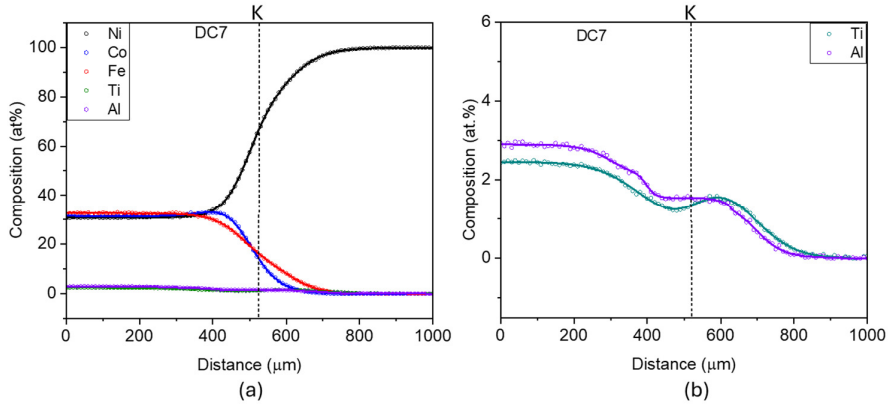

Fig. 6 (a) Diffusion profile developed in Ni – (NiCoFe)$_{94}$Ti$_3$Al$_3$ diffusion couple after 1200°C and 25 h (b) Al and Ti profile characteristics are highlighted. Location of the Kirkendall marker plane (K) is indicated by the dashed line.

Table 5 (a) Estimated tracer diffusion coefficients at the Kirkendall marker plane in the quinary Ni-Co-Fe-Al-Ti system at 1200°C. The data estimated in the ternary system at cross 1 composition (refer to Table 3) are also listed for comparison.

| Elements | Composition (at.%) (K-plane of Ni-Co-Fe-Cr-Al) | $D_i^*$ (x10$^{-15}$ m$^2$/s) | Composition (at.%) (Ternary cross 1 of Ni-Al-Ti) | $D_i^*$ (x10$^{-15}$ m$^2$/s) |
|---|---|---|---|---|
| Ni | 64.1 | 21.2±4.5 | 97.5 | 11 [53] |
| Co | 15.3 | 22.3±3.5 | | |
| Fe | 17.6 | 47.2±7.1 | | |
| Ti | 1.4 | 51.0±7.7 | 1.4 | 40.1±6 |
| Al | 1.6 | 70.1±10.2 | 1.1 | 43.0±6.5 |

This diffusion couple is strategically designed for coupling pure Ni with (NiCoFe)$_{94}$Al$_3$Ti$_3$ such that Co, Fe, Al and Ti have zero composition in the pure Ni end member, making it easier to calculate the diffusivities with less error, which is explained in detail in Ref. [4]. In such a scenario, Eq. 3b reduces to $V_m J_i = -\frac{1}{2t}\left[N_i^+ \int_{x^-}^{x_K} Y_i dx\right]$ since $N_i^+$ is zero. On the other hand, for Ni, $N_i^-$ and $N_i^+$ are not zero. Therefore, because of the minus sign in the bracketed term, a small error on both sides can increase the error on the estimated data significantly. Since, again, the interdiffusion process is mainly controlled by the composition of minor elements, this design strategy of diffusion couple is ideal. A discussion of the advantages of such a design of diffusion couples and error analysis is discussed in Ref. [4]. The diffusion couple produced at 1200°C after annealing for 25 h is shown in Fig. 6a. The composition profiles of Al and Ti are highlighted in Fig. 6b. The location of the Kirkendall marker plane (K) is identified with the dashed line at which the diffusion coefficients are estimated. The intrinsic fluxes are first calculated at this plane. Following, these are related directly to the tracer diffusion coefficients by replacing Eq. 5 in 3a. The tracer diffusion coefficients are then calculated by utilizing the thermodynamic factors calculated by extracting the activity data in this system from the ThermoCalc database TCNI9 [54]. These are listed in Table 5. We have compared the tracer diffusion coefficients estimated in the Ni-Al-Ti (taken from Table 3). The trend of diffusivities $D^*_{Ni} < D^*_{Ti} < D^*_{Al}$ in the ternary and multicomponent system, it is the same. Moreover, this trend of diffusivities $D^*_{Ni} \approx D^*_{Co} < D^*_{Fe}$ is the same as the tracer diffusion coefficients calculated in another ternary system Ni-Co-Fe [4] and Ni-Co-Fe-Cr [15] systems. The trend of diffusivities $D^*_{Ni} \approx D^*_{Co} < D^*_{Fe} < D^*_{Al}$ is the same as the tracer diffusion coefficients estimated in another quinary system Ni-Co-Fe-Cr-Al [17]. Even the self and impurity diffusion coefficients in pure Ni reported in literature also follows the same trend of $D^*_{Ni} \approx D^*_{Co} < D^*_{Fe} < D^*_{Ti} \approx D^*_{Al}$. At 1200°C, these values are reported as $D^*_{Ni} = 15.3 \times 10^{15}$ [53], $D^*_{Co}(D^{imp}_{Co(Ni)}) = 16.5 \times 10^{15}$ [55], $D^*_{Fe}(D^{imp}_{Fe(Ni)}) = 36.7 \times 10^{15}$ [56], $D^*_{Ti}(D^{imp}_{Ti(Ni)}) = 61.8 \times 10^{15}$ [57] or $65.2 \times 10^{15}$ [58] and $D^*_{Al}(D^{imp}_{Al(Ni)}) = 6.02 \times 10^{15}$ [59] or $58.6 \times 10^{15}$ [61] (all in $m^2/s$). The diffusion coefficients (estimated at the K plane) in the quinary system are found to be slightly higher than the diffusivities in the ternary systems at the equivalent composition of Ti and Al (as shown in Table 5). This is very similar to the trend of PB interdiffusion coefficient of Al-Ti in these ternary and quinary systems (as already shown in Fig. 5 [b]. These details indicate the consistency of data estimated in this study.

The intrinsic and interdiffusion coefficients are then calculated from the tracer diffusion coefficients following Eq. 5 and 4, respectively, as listed in Table 6. The differences (values and/or sign) of the main intrinsic and interdiffusion coefficients are not significant. However, a certain cross intrinsic and interdiffusion coefficients have very different values and sometimes even signs. For example, $D^{Ni}_{CoFe}$ and $\widetilde{D}^{Ni}_{CoFe}$ have very different values although the signs are the same. Diffusional interactions of elements have mostly been discussed until now (over the decades), considering the interdiffusion coefficients only. If we discuss the same in this system, the diffusional interactions of Co with Fe considering the interdiffusion coefficient ($\widetilde{D}_{CoFe}^{Ni}$, calculated considering the vacancy wind effect) in comparison to the main interdiffusion coefficient of Co ($\widetilde{D}_{CoCo}^{Ni}$) is $\frac{\widetilde{D}_{CoFe}^{Ni}}{\widetilde{D}_{CoCo}^{Ni}} = -\frac{8.5}{19.6} = -0.43$. However, the same considering the intrinsic diffusion coefficients is correctly identified as $\frac{D_{CoFe}^{Ni}}{D_{CoCo}^{Ni}} = -\frac{1.7}{20.7} = -0.08$. It should be noted here that the interdiffusion coefficients are kind of average of $n$ intrinsic diffusion coefficients such that we have (refer to Eq. 4) $\widetilde{D}_{CoFe}^{Ni} = (1 - N_{Co})D_{CoFe}^{Ni} - N_{Co}(D_{NiFe}^{Ni} + D_{FeFe}^{Ni} + D_{TiFe}^{Ni} + D_{AlFe}^{Ni}) = [(1 - 0.153)(-1.7) - 0.153(-29.2 + 61.4 + 7.7 + 6.2)] \times 10^{-15} = -8.5 \times 10^{-15} \, m^2/s$ compared to the intrinsic diffusion coefficient $D_{CoFe}^{Ni} = -1.7 \, m^2/s$. A higher value of the interdiffusion coefficient $\widetilde{D}_{CoFe}^{Ni}$ resulted from a much higher value of other intrinsic diffusion coefficients. Therefore, it is clear that only the intrinsic diffusion coefficient signifies the actual diffusion rate or diffusional interaction of an element. Any discussion considering only the interdiffusion coefficient can be misleading unless the composition of the element of interest ($N_{Co}$ in this case) is significantly low to have $\widetilde{D}_{CoCo}^{Ni} \approx D_{CoFe}^{Ni}$ [4]. Another problem of discussing the diffusional interactions with the interdiffusion coefficient instead of the intrinsic diffusion coefficient can be understood from the values of $D_{CoAl}^{Ni} = 7.4 \times 10^{-15}$ and $\widetilde{D}_{CoAl}^{Ni} = -9.4 \times 10^{-15} \, m^2/s$ (considering VWE). While $D_{CoAl}^{Ni}$ has a significant value (compared to $D_{CoCo}^{Ni}$) with a positive sign indicates significant positive diffusional interactions. On the other hand, even higher values but a negative sign of $\widetilde{D}_{CoAl}^{Ni}$ wrongly indicates a very significant negative diffusional interaction if only the interdiffusion coefficient values are calculated for such a discussion. Such a discussion in other systems can be found in Ref. [4, 17, 50]. It is important to highlight such issues in different systems to consider the importance of estimating/calculating intrinsic diffusion coefficients, which are hardly calculated/estimated even in ternary systems.

Table 6 (a) Thermodynamic factors calculated from the activity data available in ThermoClac database TCNI9, (b) intrinsic and (c) interdiffusion coefficients calculated neglecting and considering the vacancy wind effects from the tracer diffusion coefficients estimated at the Kirkendall marker plane (as listed in Table 5).

| $\phi_{ij}^{Ni}$ | | $D_{ij}^n$ | Without VWE ($\times 10^{-15}\,m^2/s$) | With VWE ($\times 10^{-15}\,m^2/s$) | $(1+W_{ij})$ | $\widetilde{D}_{ij}^n$ | Without VWE ($\times 10^{-15}\,m^2/s$) | With VWE ($\times 10^{-15}\,m^2/s$) | $1+W_i$ |
|---|---|---|---|---|---|---|---|---|---|
| $\phi_{NiCo}^{Ni}$ | -0.26 | $D_{NiCo}^{Ni}$ | -23.1 | -22.6 | 0.98 | | | | |
| $\phi_{NiFe}^{Ni}$ | -0.39 | $D_{NiFe}^{Ni}$ | -30.1 | -29.2 | 0.97 | | | | |
| $\phi_{NiTi}^{Ni}$ | -0.07 | $D_{NiTi}^{Ni}$ | -67.9 | -65.4 | 0.96 | | | | |
| $\phi_{NiAl}^{Ni}$ | -0.06 | $D_{NiAl}^{Ni}$ | -51.0 | -38.7 | 0.76 | | | | |
| $\phi_{CoCo}^{Ni}$ | 0.92 | $D_{CoCo}^{Ni}$ | 20.5 | 20.7 | 1.01 | $\widetilde{D}_{CoCo}^{Ni}$ | 19.5 | 19.6 | 1.01 |
| $\phi_{CoFe}^{Ni}$ | -0.1 | $D_{CoFe}^{Ni}$ | -1.9 | -1.7 | 0.89 | $\widetilde{D}_{CoFe}^{Ni}$ | -8.5 | -8.5 | 1.01 |
| $\phi_{CoTi}^{Ni}$ | 0.04 | $D_{CoTi}^{Ni}$ | 9.75 | 9.32 | 0.96 | $\widetilde{D}_{CoTi}^{Ni}$ | 8.8 | 8.2 | 0.93 |
| $\phi_{CoAl}^{Ni}$ | 0.02 | $D_{CoAl}^{Ni}$ | 4.3 | 7.4 | 1.73 | $\widetilde{D}_{CoAl}^{Ni}$ | -8.7 | -9.4 | 1.08 |
| $\phi_{FeFe}^{Ni}$ | 1.29 | $D_{FeFe}^{Ni}$ | 60.9 | 61.4 | 1.01 | $\widetilde{D}_{FeFe}^{Ni}$ | 53.4 | 53.6 | 1.00 |
| $\phi_{FeCo}^{Ni}$ | 0.002 | $D_{FeCo}^{Ni}$ | 0.1 | 0.4 | 3.9 | $\widetilde{D}_{FeCo}^{Ni}$ | -1.0 | -0.91 | 0.91 |
| $\phi_{FeTi}^{Ni}$ | 0.1 | $D_{FeTi}^{Ni}$ | 59.3 | 60.8 | 1.03 | $\widetilde{D}_{FeTi}^{Ni}$ | 44.2 | 44.9 | 1.02 |
| $\phi_{FeAl}^{Ni}$ | 0.06 | $D_{FeAl}^{Ni}$ | 31.1 | 38.6 | 1.24 | $\widetilde{D}_{FeAl}^{Ni}$ | 16.2 | 19.4 | 1.20 |
| $\phi_{TiTi}^{Ni}$ | 1.32 | $D_{TiTi}^{Ni}$ | 67.3 | 67.3 | 1.0 | $\widetilde{D}_{TiTi}^{Ni}$ | 66.1 | 66.2 | 1.00 |
| $\phi_{TiCo}^{Ni}$ | 1.11 | $D_{TiCo}^{Ni}$ | 5.2 | 5.2 | 1.0 | $\widetilde{D}_{TiCo}^{Ni}$ | 5.1 | 5.1 | 1.00 |
| $\phi_{TiFe}^{Ni}$ | 1.9 | $D_{TiFe}^{Ni}$ | 7.7 | 7.7 | 1.01 | $\widetilde{D}_{TiFe}^{Ni}$ | 7.1 | 7.1 | 1.00 |
| $\phi_{TiAl}^{Ni}$ | 0.27 | $D_{TiAl}^{Ni}$ | 12.0 | 12.7 | 1.06 | $\widetilde{D}_{TiAl}^{Ni}$ | 10.8 | 11.1 | 1.03 |
| $\phi_{AlAl}^{Ni}$ | 1.26 | $D_{AlAl}^{Ni}$ | 88.3 | 89.3 | 1.01 | $\widetilde{D}_{AlAl}^{Ni}$ | 86.9 | 87.6 | 1.01 |
| $\phi_{AlCo}^{Ni}$ | 0.51 | $D_{AlCo}^{Ni}$ | 3.74 | 3.78 | 1.01 | $\widetilde{D}_{AlCo}^{Ni}$ | 3.6 | 3.67 | 1.02 |
| $\phi_{AlFe}^{Ni}$ | 0.96 | $D_{AlFe}^{Ni}$ | 6.1 | 6.2 | 1.01 | $\widetilde{D}_{AlFe}^{Ni}$ | 5.4 | 5.4 | 1.00 |
| $\phi_{AlTi}^{Ni}$ | 0.22 | $D_{AlTi}^{Ni}$ | 17.6 | 17.8 | 1.01 | $\widetilde{D}_{AlTi}^{Ni}$ | 16.2 | 16.4 | 1.01 |
| (a) | | (b) | | | | (c) | | | |

## 3.2 Constraint-enhanced PINN method for extraction of composition-dependent diffusivities

One of the major drawbacks of experimental diffusion couple methods (without the use of radioisotopes) is that we can estimate the diffusion coefficients only at certain compositions, such as the composition of interaction or at the Kirkendall marker plane, as described in this article. Therefore, the PINN numerical inverse method is practiced for extracting composition-dependent diffusion coefficients. The equation scheme and methodology are already explained in Section 2. A few important points to be further noted here. Most numerical inverse or CALPHAD-based methods established so far do not account for the vacancy wind effect, whereas our approach explicitly includes it for more accurate calculations. Moreover, our method does not require prior knowledge of diffusivities and converges reliably even from random initial guesses — an important feature given the high-dimensional parameter space. In the quinary system, up to 11 polynomial coefficients are needed for each element (55 in total), yet the optimization consistently reaches a physically meaningful solution. As already explained in Ref. [50], a moderate-sized neural network architecture is employed with layer-wise training, and Gaussian Process (GP)-based Bayesian optimisation is used to tune the hyperparameters and enhance convergence performance.

We have noticed that optimization cannot rely on composition profiles alone because several different sets of tracer diffusion coefficients can reproduce the same profile, leading to non-unique or unphysical solutions. For example, as shown in Fig. 7a, optimization with the composition profile of the ternary diffusion couple in the Ni-Al-Ti shows an excellent match. However, without the use of estimated tracer diffusion coefficients as the equality constraint, the extracted tracer diffusion coefficients are not reliable, as shown in Fig. 7b. The extracted tracer diffusion coefficient of Ni is higher than the tracer diffusion coefficients of Al and Ti in one composition range and different in another is illogical considering the expected relative mobilities of elements, as already discussed. Moreover, tracer diffusion coefficients of Al and Ti are almost equal in low Ni content, but significantly higher in high Ni content, which is also illogical since diffusion coefficients are not expected to vary so drastically in such a small composition range in a solid solution. Without using any experimentally estimated tracer diffusion coefficients, the optimization may produce any random data, ensuring the matching of the diffusion profile, since several combinations of values of tracer diffusion coefficients can match the profile. We have already measured tracer diffusion coefficients at three different close compositions along this ternary diffusion profile. For understanding the stability and performance of the outcome after optimization, we decided to use the impurity and self-diffusion coefficient in the pure end member and tracer diffusion coefficients at the cross composition 2 as equality constraints.

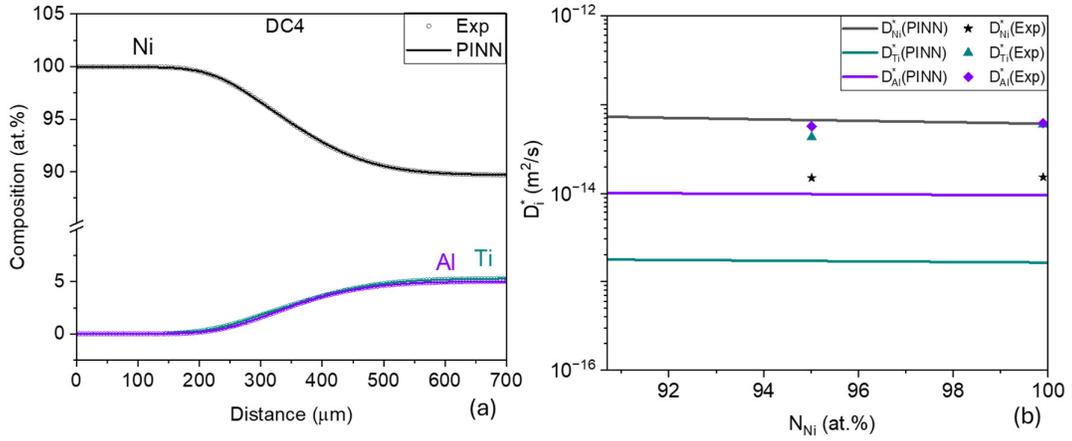

Fig. 7 (a) Comparison of the PINN-optimized diffusion profile with the experimentally measured ternary diffusion profile after fitting and smoothing (see Fig. 2a). (b) Tracer diffusion coefficients extracted from PINN optimization without equality constraints compared with experimentally measured tracer diffusivities and impurity diffusion coefficients in Ni, highlighting the unreliability of profile-only optimization.

The self and impurity diffusion coefficients in Ni estimated experimentally in pure metals are generally found in the literature. Therefore, these data can play an important role by using them as an equality constraint in the pure end member side of the diffusion couple. Moreover, these are also useful as a reference for the data estimated in ternary and multicomponent systems, especially in solid solutions, for understanding the reliability of the data estimated or indicating the systematic variation with composition. For example, we have noticed that the overall value of the diffusion coefficients may change, but relative mobilities follow the same trend in most of the solid solutions [4,17, 50], for example $D^*_{Ni} < D^*_{Ti} \approx D^*_{Al}$ in the Ni-Al-Ti system and $D^*_{Ni} \approx D^*_{Co} < D^*_{Fe} < D^*_{Ti} \approx D^*_{Al}$ in the quinary Ni-Co-Fer-Al-Ti quinary system. The impurity and self-diffusion coefficients available in literature and considered in this study are listed in Table 7. To understand, if there is any difference in data estimated by us compared to the data available in the literature, we conducted Ni-Al and Ni-Ti binary diffusion couple experiments at 1200°C as shown in Fig. 8. The interdiffusion coefficients are calculated following the Eq. 1b and 6a. The impurity diffusion coefficients, $D^{imp}_{Ti(Ni)}$ and $D^{imp}_{Al(Ni)}$ are calculated by extending the interdiffusion coefficients to zero Ti and Al and compared with the data available in literature to find an excellent match. The interdiffusion coefficient is almost equal to the intrinsic diffusion coefficients of minor elements in this low concentration range of Ti and Al $(N_{i(Ti/Al)} \ll N_{Ni})$ since we have $\widetilde{D} = N_{Ni}D_{i(Ti/Al)} + N_{i(Ti/Al)}D_{Ni} \approx D_{i(Ti/Al)}$. The increase in interdiffusion coefficient with the increase in composition of minor elements indicates the increase in intrinsic diffusion coefficients of these elements. As already discussed, a similar trend is also found in the Ni-Al-Ti system (refer to Fig. 3).

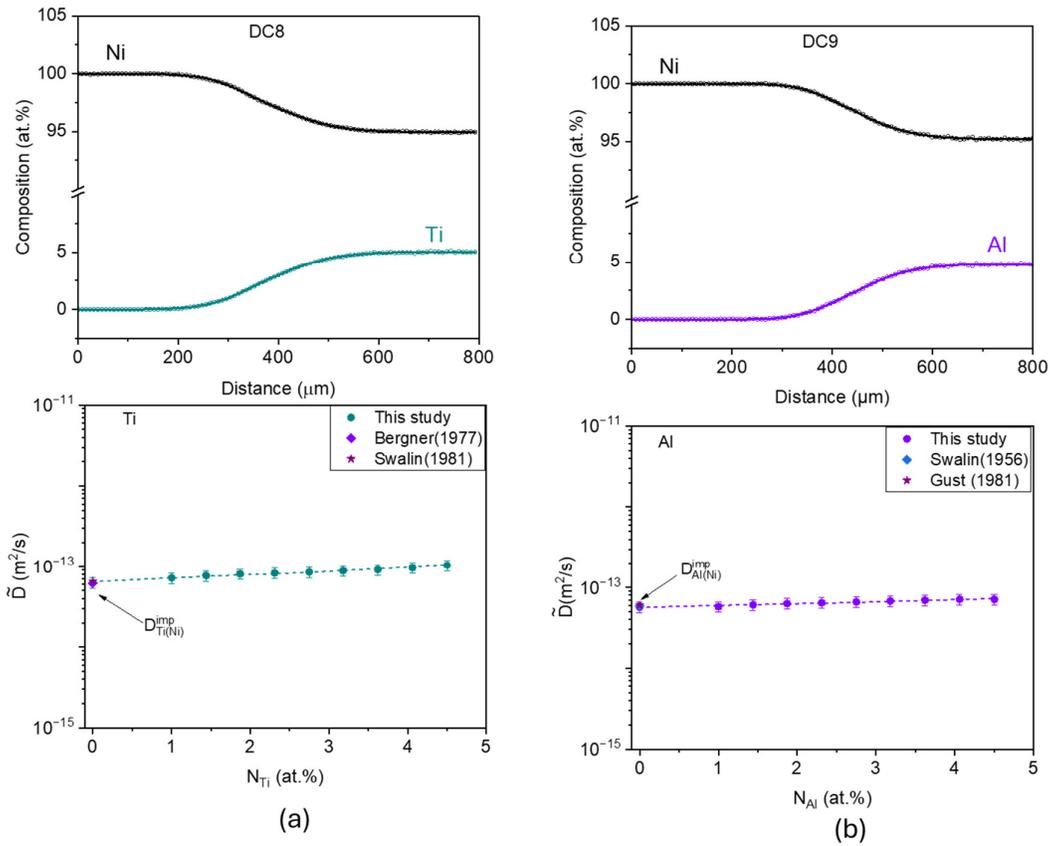

Fig. 8 (a) Ni-Ti and (b) Ni-Al binary diffusion couple experiments at 1200 °C annealed for 16 h. The estimated interdiffusion coefficients are extended to zero Ni to calculate the impurity diffusion coefficients and compare them with the data available in the literature.

Table 7 (a) Self and impurity diffusion coefficients of Ni, Al and Ti in pure Ni at different temperatures.

|  | 1050°C (x10$^{-15}$ m$^2$/s) | 1100°C (x10$^{-15}$ m$^2$/s) | 1150°C (x10$^{-15}$ m$^2$/s) | 1200°C (x10$^{-15}$ m$^2$/s) |
|---|---|---|---|---|
| $D_{Ni(Ni)}^{\text{self}}$ [53] | 1.1 | 2.9 | 6.8 | 15.3 |
| $D_{Ti(Ni)}^{\text{imp}}$ [57] | 5.3 | 12.9 | 29.0 | 61.8<br>65.2 [58]<br>62.4±0.9 (This Study) |
| $D_{Al(Ni)}^{\text{imp}}$ [59] | 5.4 | 12.8 | 28.5 | 60.2<br>58.6 [58]<br>57.3±0.8 (This Study) |

As a next step of optimization, the self and impurity diffusion coefficients in the pure Ni and data estimated at the cross composition 2 (refer to Table 2) are used as the equality constraint. Note that the impurity diffusion coefficients cannot be utilized as equality constraints exactly at zero composition of Al and Ni in the absence of diffusion profiles of the elements. These are, therefore, considered at 0.5 at.% composition since diffusion coefficients at these compositions can be considered similar. As shown in Fig. 9, again, a very nice match with the smoothed experimental diffusion profile can be seen. Additionally, the tracer diffusion coefficient variations are now reliable. Thus, incorporating tracer diffusivities at intersection points together with self- and impurity-diffusion coefficients near the pure Ni end effectively anchors the optimization and removes spurious solutions, ensuring that the resulting diffusivity functions follow physically meaningful trends.

The optimization parameters at different temperatures are listed in Table 8, considering the expressions for the tracer diffusion coefficients of elements in reference to Eq. 10 as

$$D_i^* = \exp\left(\theta_0^i + \theta_{Ti}^{1,i} N_{Ti} + \theta_{Al}^{1,i} N_{Al} + \theta_{Al,Ti}^{2,i} N_{Ti} N_{Al}\right) \quad (23)$$

Considering optimization parameter, the tracer diffusion coefficients of Al and Ti are calculated at cross composition 2 (refer to Table 2) and compared with the experimentally estimated tracer diffusion coefficients and activation energy to find an excellent match, as shown in Fig. 10. This is expected since the optimization parameters are obtained using tracer diffusion coefficients experimentally as equality constraints. This is a very important step for extracting reliable data following any numerical optimization, which has hardly been practiced until now. The calculated intrinsic diffusion coefficients at different cross compositions are compared in Fig. 3 along with experimentally estimated intrinsic diffusion coefficients. We have used self and impurity diffusion coefficients in pure Ni and tracer diffusion coefficient values at the cross composition 2 only as the equality constraints. One can also use additional data estimated at other locations (cross-composition 1 and 3) as the equality constraints. However, the data extracted from other compositions already matches reasonably well, and this step with additional constraints is not required. This is system-specific and may be necessary for another system, especially if the composition range is higher with higher variation of diffusion coefficients.

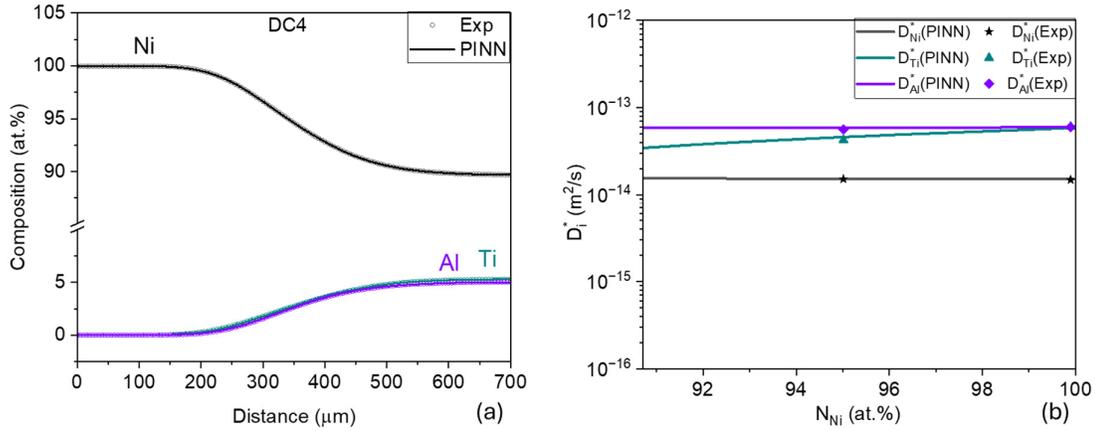

Fig. 9 (a) Matching the PINN optimized diffusion profile with experimentally developed diffusion profile after fitting and smoothing (refer to Fig. 2a) (b) comparison of extracted tracer diffusion coefficients after PINN with experimentally estimated tracer diffusion coefficients at cross composition 2 and impurity diffusion coefficients in Ni (refer to Table 7).

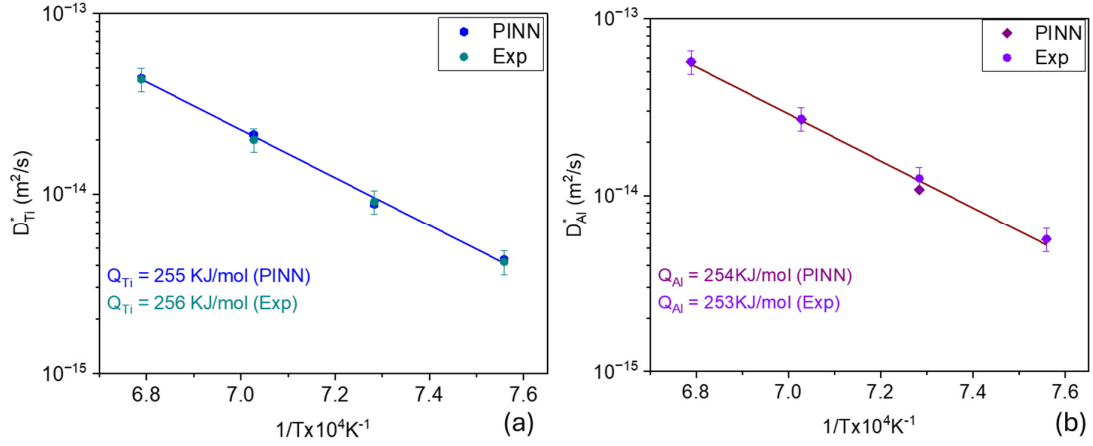

Fig. 10 Comparison of tracer diffusion coefficients and activation energy of (a) Ti and (b) Al estimated from experimental method and extracted from PINN method at cross composition 2 (refer to Table 2) at different temperatures.

Table 8 Optimization parameters for Ni, Ti and Al at different temperatures in the Ni-Al-Ti system.

| $D^*_{Ni}$ | 1050°C | 1100°C | 1150°C | 1200°C |
|---|---|---|---|---|
| $\theta_0^{Ni}$ | 1.38 | 2.35 | 3.2 | 3.94 |
| $\theta_{Ti}^{1,Ni}$ | -0.04 | -0.02 | -0.17 | 0.97 |
| $\theta_{Al}^{1,Ni}$ | -0.01 | -0.014 | -0.16 | 0.89 |
| $\theta_{Al,Ti}^{2,Ni}$ | 0.86 | 0.75 | 0.29 | 0.96 |
| $D^*_{Ti}$ | | | | |
| $\theta_0^{Ti}$ | 3.07 | 3.71 | 4.67 | 5.17 |
| $\theta_{Ti}^{1,Ti}$ | -6.43 | -1.1 | -5.4 | -2.79 |
| $\theta_{Al}^{1,Ti}$ | -6.42 | -1.1 | -5.4 | 4.6 |
| $\theta_{Al,Ti}^{2,Ti}$ | -5.37 | -1.0 | -5.4 | 4.4 |
| $D^*_{Al}$ | | | | |
| $\theta_0^{Al}$ | 2.95 | 3.81 | 4.67 | 5.35 |
| $\theta_{Ti}^{1,Al}$ | 1.02 | -1.05 | -1.4 | 0.38 |
| $\theta_{Al}^{1,Al}$ | 0.98 | -1.05 | -1.12 | 1.47 |
| $\theta_{Al,Ti}^{2,Al}$ | 1.2 | -0.98 | -1.13 | 1.31 |

To understand the quality of the optimized parameters, we calculated the tracer diffusion coefficients of Al and Ti following Eq. 23 along the PB 5Ti-5Al (95 at.%Ni constant) diffusion profile (refer to Fig. 1b) utilizing the optimized parameters as listed in

Table 8. The intrinsic diffusion coefficients of Al and Ti are then calculated following Eq. 8b and c. Subsequently, the PB interdiffusion coefficients are calculated following Eq. 8a. These are then compared with the directly estimated PB interdiffusion coefficient from the diffusion profile following Eq. 7d. We have used tracer diffusion coefficients at one composition only along the PB interdiffusion path (at the intersecting composition with the ternary diffusion profile). Still, although it remains constant, we have an excellent match in the interdiffusion coefficient over the whole composition range. To understand the extendibility of the optimized parameters to the lower order binary systems, the tracer diffusion coefficients are first calculated along the Ni-Ti and Ni-Al binary diffusion couples (refer to Fig. 8) following:

Ni-Ti binary system:

$$D^*_{Ni} = \exp(\theta_0^{Ni} + \theta_{Ti}^{1,Ni} N_{Ti}) \qquad (24a)$$

$$D^*_{Ti} = \exp(\theta_0^{Ti} + \theta_{Ti}^{1,Ti} N_{Ti}) \qquad (24b)$$

Ni-Al binary system:

$$D^*_{Ni} = \exp(\theta_0^{Ni} + \theta_{Al}^{1,Ni} N_{Al}) \qquad (25a)$$

$$D^*_{Al} = \exp(\theta_0^{Al} + \theta_{Al}^{1,Al} N_{Al}) \qquad (25b)$$

Subsequently, the intrinsic diffusion coefficients are calculated following Eq. 6c and d. Then, the interdiffusion coefficients are calculated following Eq. 6b to compare with directly estimated interdiffusion coefficients from the diffusion profiles. Again, a very good match is found, as shown in Fig. 11b and c. Most importantly, it correctly shows the trend of increasing interdiffusion coefficients with the increase in Ti and Al composition.

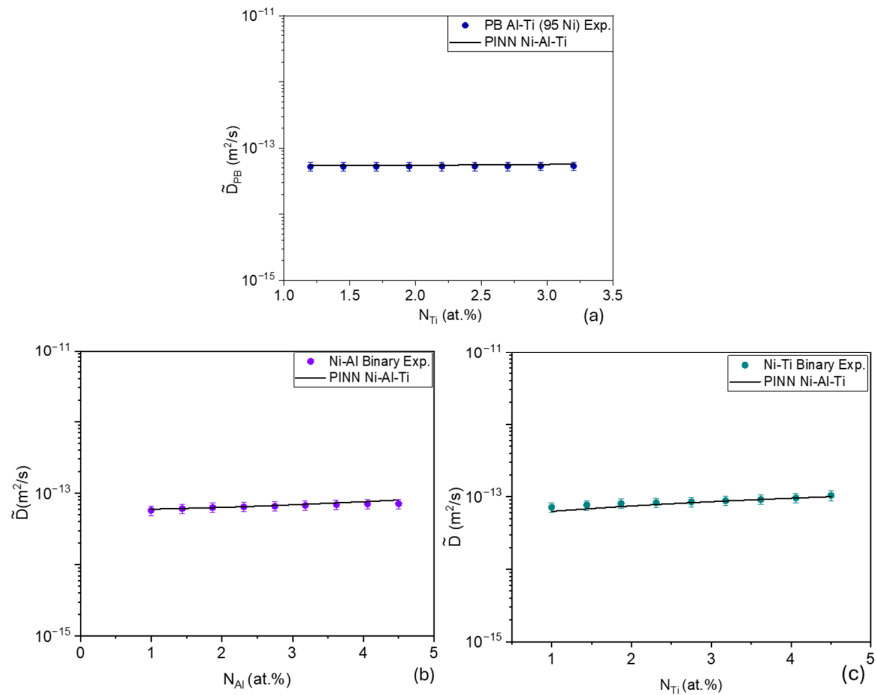

Fig. 11 Comparison of experimentally estimated interdiffusion coefficients and calculated values by extracting the optimization parameters from Ni-Al-Ti system (refer to Table 8 in (a) PB 5Al-5Ti (95 at.%Ni constant) (b) binary Ni-Al, (c) binary Ni-Ti and at 1200 ºC.

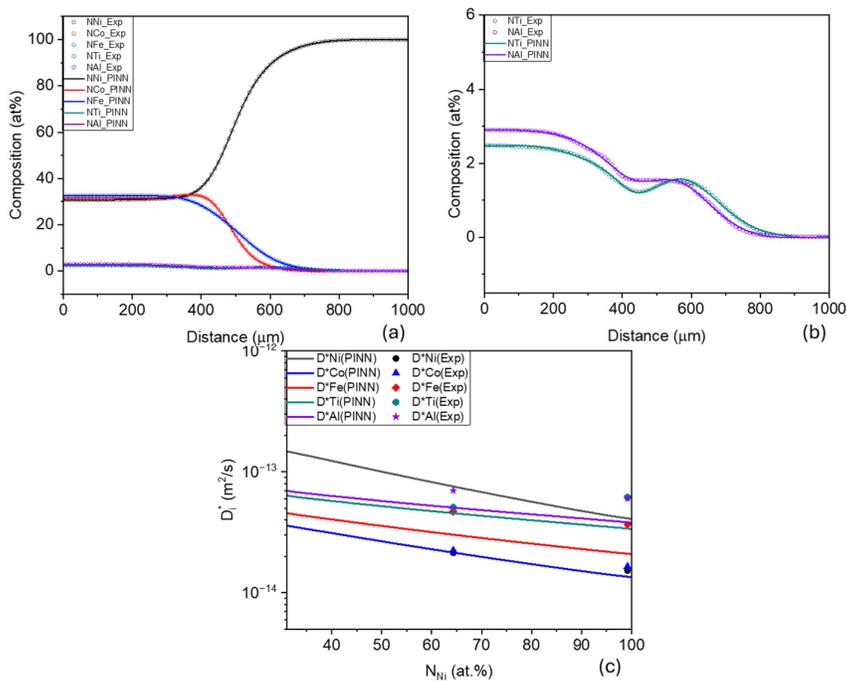

Fig. 12 (a) Comparison between experimentally produced diffusion profile of Ni – (NiCoFe)$_{94}$Ti$_3$Al$_3$ diffusion couple after 1200ºC and 25 hr and PINN optimized diffusion profile (b) comparison of Ti and Al profiles are highlighted (c) comparison between optimized tracer diffusion coefficients produced without equality constraints compared to experimentally estimated tracer diffusion coefficients at the Kirkendall marker plane in this study and impurity diffusion coefficients in Ni.

Next, we aim to establish the optimization parameters in the Ni-Co-Fe-Al-Ti quinary system. For this, we first optimized the quinary diffusion profile produced experimentally, as shown in Fig. 6. Initially, this is optimized without any tracer or impurity diffusion coefficients as equality constraints. As can be seen on Fig. 12, a nice match in the smoothed experimental diffusion profile and the optimized profile is established. However, the extracted tracer diffusion coefficients are found to be very different from the experimentally estimated diffusion coefficients at the Kirkendall marker plane and impurity diffusion coefficients in Ni. Therefore, matching with the diffusion profile does not guarantee a reliable output of tracer diffusion coefficients. In the next step, the optimization is done using tracer and impurity diffusion coefficients as equality constraints, along with matching the diffusion profile. The impurity diffusion coefficients of $D^{imp}_{Co(Ni)} = 16.5 \times 10^{-15}$ [55] and $D^{imp}_{Fe(Ni)} = 36.7 \times 10^{-15}$ [56] are taken from the literature. Self-diffusion coefficient of Ni and impurity diffusion coefficients of Ti and Al are mentioned in Table 7. It can be noticed, as shown in Fig. 13, that again an excellent match with the diffusion profile is witnessed along with reliable extraction of tracer diffusion coefficients. Impurity diffusion coefficients are used as equality constraints, almost at 0.5 at.% of Co, Fe, Ti and Al and 99.5 at.% of Ni, since these cannot be done at 100 at.% Ni without any developing diffusion profile of elements in the unaffected end member. The optimized parameters for an element *i* (=Ni, Co, Fe, Al, Ti) are listed in Table 9 in reference to Eq. 11 as

$$D_i^* = exp\big(\theta_0^i + \theta_{Co}^{1,i} N_{Co} + \theta_{Fe}^{1,i} N_{Fe} + \theta_{Ti}^{1,i} N_{Ti} + \theta_{Al}^{1,i} N_{Al} + \theta_{Co,Fe}^{2,i} N_{Co} N_{Fe} + \theta_{Co,Ti}^{2,i} N_{Co} N_{Ti} + \theta_{Co,Al}^{2,i} N_{Co} N_{Al} + \theta_{Fe,Ti}^{2,i} N_{Fe} N_{Ti} + \theta_{Fe,Al}^{2,i} N_{Fe} N_{Al} + \theta_{Ti,Al}^{2,i} N_{Ti} N_{Al}\big)$$

(26)

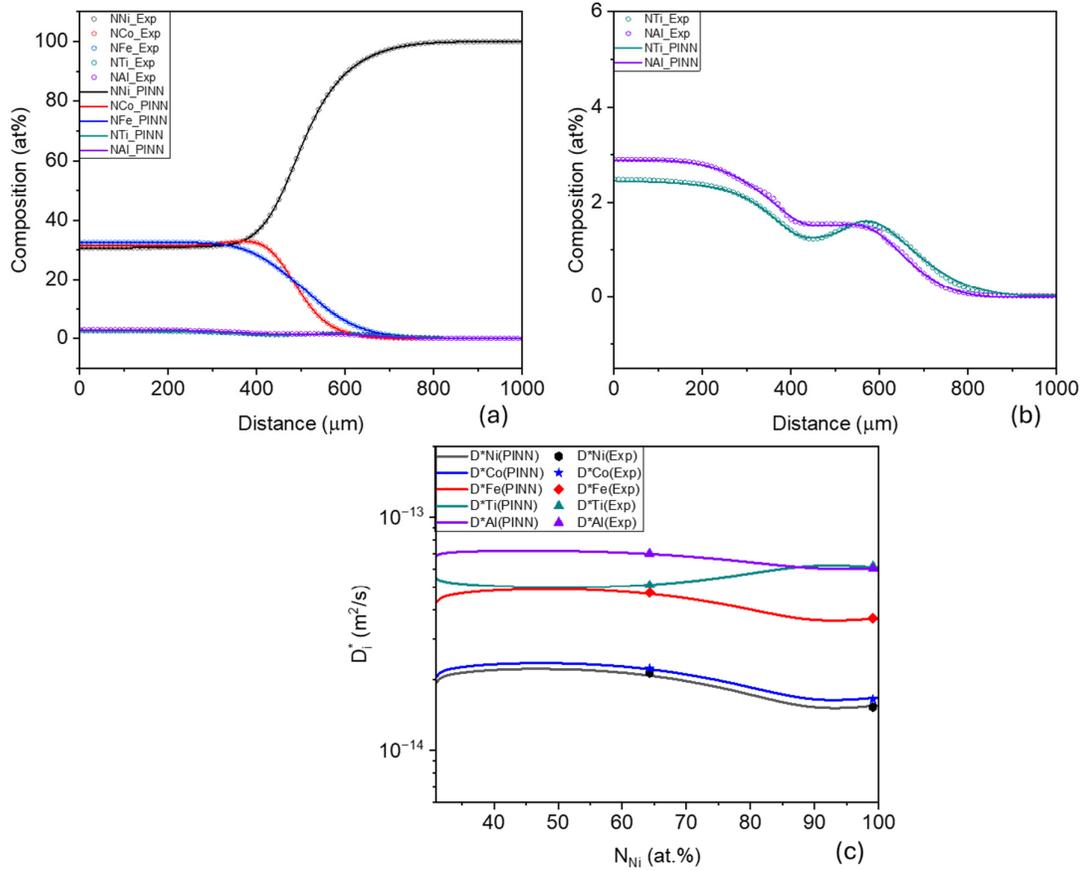

Fig. 13 (a) Comparison between experimentally produced diffusion profile of Ni – (NiCoFe)$_{94}$Ti$_3$Al$_3$ diffusion couple after 1200ºC and 25 hr and PINN optimized diffusion profile (b) comparison of Ti and Al profiles are highlighted (c) comparison between optimized tracer diffusion coefficients produced with equality constraints compared to experimentally estimated tracer diffusion coefficients at the Kirkendall marker plane in this study and impurity diffusion coefficients in Ni.

Table 9 Optimization parameters for Ni, Co, Fe, Ti and Al solid solution at 1200 °C.

| $D_{Ni}^*$ |  | $D_{Co}^*$ |  | $D_{Fe}^*$ |  | $D_{Ti}^*$ |  | $D_{Al}^*$ |  |
|---|---|---|---|---|---|---|---|---|---|
| $\theta_0^{Ni}$ | 3.96 | $\theta_0^{Co}$ | 4.03 | $\theta_0^{Fe}$ | 4.84 | $\theta_0^{Ti}$ | 5.36 | $\theta_0^{Al}$ | 5.33 |
| $\theta_{Co}^{1,Ni}$ | 1.36 | $\theta_{Co}^{1,Co}$ | 1.31 | $\theta_{Co}^{1,Fe}$ | 1.22 | $\theta_{Co}^{1,Ti}$ | 0.41 | $\theta_{Co}^{1,Al}$ | 1.02 |
| $\theta_{Fe}^{1,Ni}$ | 1.35 | $\theta_{Fe}^{1,Co}$ | 1.30 | $\theta_{Fe}^{1,Fe}$ | 1.22 | $\theta_{Fe}^{1,Ti}$ | 0.46 | $\theta_{Fe}^{1,Al}$ | 1.03 |
| $\theta_{Ti}^{1,Ni}$ | 0.83 | $\theta_{Ti}^{1,Co}$ | 0.87 | $\theta_{Ti}^{1,Fe}$ | 0.91 | $\theta_{Ti}^{1,Ti}$ | 1.58 | $\theta_{Ti}^{1,Al}$ | 1.02 |
| $\theta_{Al}^{1,Ni}$ | 0.85 | $\theta_{Al}^{1,Co}$ | 0.88 | $\theta_{Al}^{1,Fe}$ | 0.92 | $\theta_{Al}^{1,Ti}$ | 1.63 | $\theta_{Al}^{1,Al}$ | 1.04 |
| $\theta_{Co,Fe}^{2,Ni}$ | 1.39 | $\theta_{Co,Fe}^{2,Co}$ | 1.32 | $\theta_{Co,Fe}^{2,Fe}$ | 1.23 | $\theta_{Co,Fe}^{2,Ti}$ | 0.39 | $\theta_{Co,Fe}^{2,Al}$ | 1.07 |
| $\theta_{Co,Ti}^{2,Ni}$ | 0.68 | $\theta_{Co,Ti}^{2,Co}$ | 0.73 | $\theta_{Co,Ti}^{2,Fe}$ | 0.77 | $\theta_{Co,Ti}^{2,Ti}$ | 1.57 | $\theta_{Co,Ti}^{2,Al}$ | 0.97 |
| $\theta_{Co,Al}^{2,Ni}$ | 0.71 | $\theta_{Co,Al}^{2,Co}$ | 0.75 | $\theta_{Co,Al}^{2,Fe}$ | 0.79 | $\theta_{Co,Al}^{2,Ti}$ | 1.54 | $\theta_{Co,Al}^{2,Al}$ | 0.97 |
| $\theta_{Fe,Ti}^{2,Ni}$ | 0.68 | $\theta_{Fe,Ti}^{2,Co}$ | 0.72 | $\theta_{Fe,Ti}^{2,Fe}$ | 0.77 | $\theta_{Fe,Ti}^{2,Ti}$ | 1.58 | $\theta_{Fe,Ti}^{2,Al}$ | 0.97 |
| $\theta_{Fe,Al}^{2,Ni}$ | 0.71 | $\theta_{Fe,Al}^{2,Co}$ | 0.74 | $\theta_{Fe,Al}^{2,Fe}$ | 0.79 | $\theta_{Fe,Al}^{2,Ti}$ | 1.56 | $\theta_{Fe,Al}^{2,Al}$ | 0.96 |
| $\theta_{Ti,Al}^{2,Ni}$ | 1.21 | $\theta_{Ti,Al}^{2,Co}$ | 1.16 | $\theta_{Ti,Al}^{2,Fe}$ | 1.12 | $\theta_{Ti,Al}^{2,Ti}$ | 0.53 | $\theta_{Ti,Al}^{2,Al}$ | 1.01 |

To check the extendibility of the optimized parameters to the same or lower order diffusion profiles utilizing the optimizing parameters in this quinary system, we first calculate the interdiffusion coefficients in the binary Ni-Ti and Ni-Al systems. The calculation method following Eq. 24 and 25 is already explained in this quinary system established from only a single diffusion profile, we extract the tracer diffusion coefficients along the Ni-Ti and Ni-Al binary diffusion profiles, as shown in Fig. 8. Following, the interdiffusion coefficients estimated directly from these profiles and calculated from optimized parameters are compared in Fig. 14a and b. A very good match is evident. It also shows the trend of increasing interdiffusion coefficients slightly with the increase in composition of Ti or Al. Subsequently, we have compared the PB interdiffusion coefficient calculated from 5Ti-5Al (95 at.%Ni) ternary PB diffusion profile, following the calculation methodology already explained. Again, a very good match is clear, as shown in Fig. 14c. Next, we calculate the PB diffusion coefficient by calculating the PB interdiffusion coefficient along the PB 5Ti-5Al (95 at.% Ni+Co+Fe) diffusion profile (refer to Fig. 5a) by calculating the tracer diffusion coefficients of Ti and Al from the optimized parameters. Note that the composition range of both PB profiles (ternary and quinary) is beyond the composition range of the quinary diffusion profile. Although the match is excellent with the ternary PB diffusion profile, a similar level of match is not found with the quinary PB diffusion profile. However, the difference is not very significant. Such a level of (or even higher) difference is frequently reported in experimental results. Moreover, this is not so uncommon especially when different types of diffusion couples are produced for comparison [60] or even the typical error range mentioned in radiotracer experiments [61]. Still, it indicates that optimized parameters for calculating the tracer diffusion coefficients should not be used beyond the composition range considered for optimization. However, the difference in this system can be within the range of experimental errors.

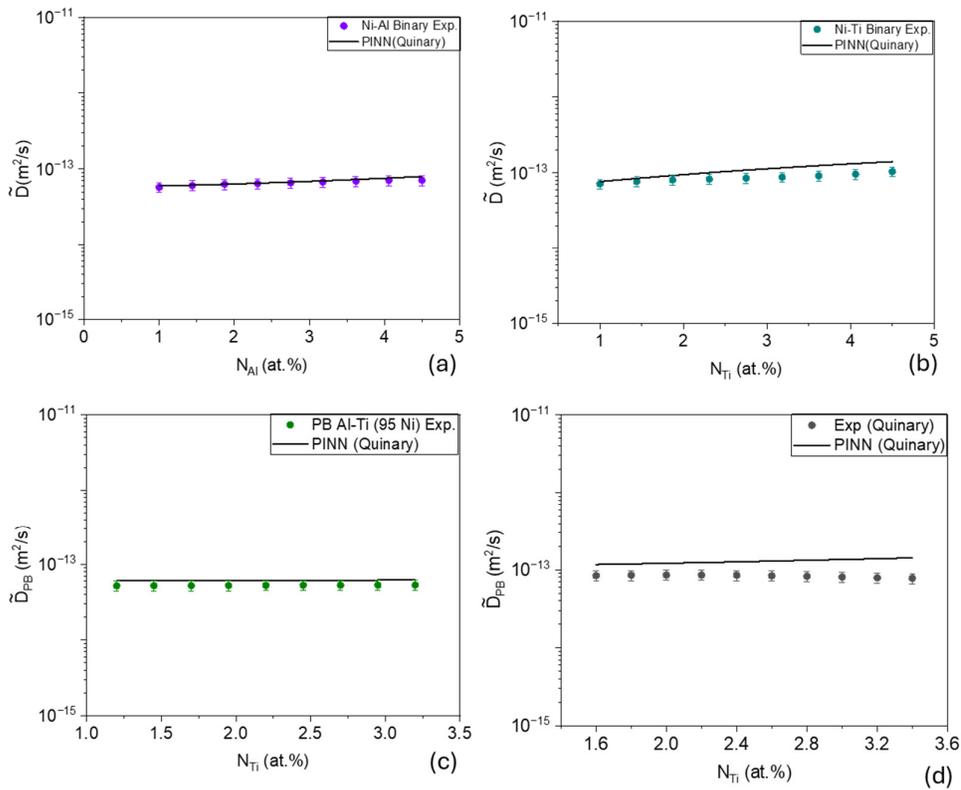

Fig. 14 Comparison of experimentally estimated interdiffusion coefficients and calculated values by extracting the optimization parameters from quinary Ni-Co-Fe-Al-Ti system (refer to Table 9 in (a) binary Ni-Al, (b) binary Ni-Ti and (c) PB 5 Al-Ti (95 at.% Ni constant) (d) PB 5 Al-Ti (95 at.% Ni+Co+Fe constant) at 1200 ºC.

## 4. Conclusion

This study reports extensive diffusion analyses of the Ni-Al-Ti ternary and Ni-Co-Fe-Al-Ti quinary systems. First, the diffusion coefficients of all the elements are estimated at certain compositions. Following, a physics-informed neural network-based numerical inverse method is practiced for extracting composition-dependent diffusion coefficients over the whole composition range of the diffusion couples. The outcome of this study can be summarized as:

1) The possibility of producing pseudo-binary diffusion profiles is demonstrated in the Ni-Ti-Al (with Ni as constant) and Ni-Co-Fe-Al-Ti (with Ni, Co and Fe as constant) solid solution for the first time. Estimating diffusion coefficients with systematic composition variation is demonstrated by intersecting a ternary diffusion profile with Al-Ti (constant Ni) PB diffusion profiles in the Ni-Al-Ti system. A single profile estimation method is followed in the quinary Ni-Co-Fe-Al-Ti system solid solution to estimate diffusion coefficients of all the elements, for the ease of experimental practice and analysis.

2) We have shown that the discussion of diffusional interactions considering the interdiffusion coefficients instead of intrinsic diffusion coefficients can be misleading, indicating wrong diffusional interactions of certain elements, since interdiffusion coefficients are a kind of average of $n$ intrinsic diffusion coefficients.

3) A single well-designed diffusion profile, combined with equality-constrained PINN optimization, can extract reliable, physically consistent tracer diffusivities. This demonstrates that constraint-enforced inverse modelling is mandatory/essential to eliminate non-unique solutions and extract reliable and unique tracer diffusivity (fundamental mobility) data. Including experimental tracer data (at specific intersecting composition or at Kirkendall marker plane) and impurity diffusivities as equality constraints, which are shown to be necessary for the generation of a reliable mobility database. When combined with equality-constrained PINN optimization, a single well-designed diffusion profile is demonstrated to be sufficient to extract physically consistent tracer diffusivities. This confirms that constraint-enforced inverse modelling is essential for eliminating non-unique solutions and obtaining unique, physically meaningful tracer diffusivity (fundamental mobility) data. Including experimental tracer data at intersecting compositions or at the Kirkendall marker plane, together with impurity diffusivities as equality constraints, is therefore crucial for building a trustworthy and transferable mobility database.

4) High-dimensional optimization in the quinary system, involving 11 parameters per element (55 in total), is effectively addressed using the PINN framework. Bayesian optimization is used for hyperparameter tuning, improving the network's convergence and generalisation. Regularization and staged training further help avoid overfitting, and convergence is achieved reliably even from random initial guesses.

5) The extendibility of the optimized parameters for diffusion coefficients is demonstrated by comparing the experimentally estimated binary interdiffusion coefficients from Ni-Al and Ni-Ti binary diffusion profiles and pseudo-binary diffusion profiles in ternary Ni-Ti-Al and Ni-Co-Fe-Ti-Al systems. However, this extendibility is not guaranteed in other systems, which depends on the diffusion coefficients in a particular system. Still, such a possibility may be witnessed in multiple systems, which need to be explored. This will ease the generation of mobility databases in various multicomponent systems, which is still a significant challenge.

**Acknowledgement:** We acknowledge the use of EPMA in the Advanced facility for Microscopy and Micro Analysis (AFMM) at IISc. We also acknowledge the financial support from SERB, India (Project No. CRG/2021/001842).